\documentclass[final]{iasart}           

\usepackage[pdftex]{graphicx}

\usepackage[utf8]{inputenc}
\usepackage[T1]{fontenc}

\usepackage{amssymb}
\usepackage[centertags]{amsmath}

\usepackage{algorithm}
\usepackage{algorithmic}

\usepackage[below]{placeins}
\usepackage{color}

\newtheorem{property}{Property}

\DeclareMathOperator*{\argmin}{argmin}

\title{Fast computation of all pairs of geodesic distances}
\shorttitle{Fast computation of all pairs of geodesic distances}
\shortauthors{Noyel G \etal}

\author{Guillaume Noyel}
\author{Jes\'us Angulo}
\author{Dominique Jeulin}

\email{\{guillaume.noyel, jesus.angulo, dominique.jeulin\}@ensmp.fr}

\affiliation{MINES ParisTech, CMM - Centre de Morphologie
Math\'ematique, Math\'ematiques et Syst\`emes, 35 rue Saint Honor\'e - 77305
Fontainebleau cedex, France}

\abstract{Computing an array of all pairs of geodesic distances
between the pixels of an image is time consuming. In the sequel, we
    introduce new methods exploiting the redundancy of geodesic propagations
    and compare them to an existing one. We show that our method in which the
    source point of geodesic propagations is chosen
    according to its minimum number of distances to the other points, improves the
    previous method up to 32 \% and the naive method up to 50 \% in terms of
    reduction of the number of operations.
}


\begin{document}
\begin{paper}

\section{Introduction}
\label{sec:intro}

An array of all pairs of geodesic distances, between nodes of a
graph, is very useful for several applications such as clustering by
kernel methods or graph-based segmentation or data analysis.
However, computing this array of distances is time consuming. That
is the reason why two new methods, that fill in a fast way the
distances array, are presented and compared in this paper.

Our methods are available on general graphs. In this paper we
present them on images of which the pixels are considered as the
nodes of the graph and the links between the neighbors corresponds
to the edges of the graph.

In image processing, it is interesting to use the geodesic distance
between pixels in place of another distance. The array of all pairs
of geodesics distances allows to segment an image in geodesically
connected regions (i.e., geodesic balls) \citep{Noyel_ISMM_2007}.
All pairs of geodesic distances are also useful on defining adaptive
neighborhoods of filters used for edge-preserving smoothing
\citep{Lerallut_2007,Bertelli_2007,Grazzini_2009}. Nonlinear
dimensionality reduction techniques are mostly based on
multidimensional scaling on a Gram matrix of distances between the
pairs of variables. Particularly interesting for estimating the
intrinsic geometry of a data manifold is the Isometric feature
mapping \emph{Isomap} \citep{Tenenbaum_1998,Tenenbaum_2000}. After
defining a neighborhood graph of variables, Isomap calculates the
shortest path between every pairs of vertices, which is then
low-dimensional embedded via multidimensional scaling. The
application of Isomap to hyperspectral image analysis requires the
computation of all pairs of geodesic distances for a graph of all
the pixels of an image \citep{Mohan_2007}.

However, the computation of all pairs of geodesic distance in an
image is time consuming. In fact, the naive approach consists in
repeating $N$ times the algorithm to compute the geodesic distance
from each of the $N$ pixels to all others. Computing the geodesic
distance from one pixel to all others is called a geodesic
propagation. The pixel at the origin of a propagation is called the
source point and the array containing all pairs of geodesic
distances is named the distances array. Therefore, the naive
approach is of complexity $O(N \times M)$, with $M$ the complexity
of a geodesic propagation. Several algorithms for geodesic
propagations are available: the most famous is the Fast Marching
Algorithm introduced by \citet{Sethian_1996,Sethian_1999} and of
complexity $O(N \log(N))$. This algorithm consists in computing
geodesic distances in a continuous domain, using a first order
approximation, to obtain the distance in the discrete domain.
Another algorithm was developed by \citet{Soille_1991} for binary
images and for grey level images \citep{Soille_1992}. In order to
compare our results, we will use the Soille's algorithm of
``geodesic time function'' \citep{Soille_1994,Soille_2003}. Recent
implementations of Soille's algorithm for binary images are in $O(N
\log(N))$ \citep{Coeurjolly_2004}. \citet{Bertelli_2006} have
introduced a method to exploit the redundancy when several geodesic
propagations are computed. In fact, when we perform a geodesic
propagation from one pixel, all the geodesic paths from this pixel
are stored in a tree. Using this tree, we know all the geodesic
distances between any pairs of points along the geodesic path
connecting two points. This redundancy is also exploited in earlier
algorithm for computing the propagation function well known in
mathematical morphology \citep{Lantuejoul_1984}.

In order to choose the source points of the geodesic propagations,
\citet{Bertelli_2006} have proposed to select them randomly in a
spiral like order starting from the edges of the image and going to
the center. In the sequel, we test several deterministic approaches
to select the source points and we show that a method based on the
filling rate of the distances array can reduce the number of
operations up to 32 \% compared to \citet{Bertelli_2006} method.

After discussing some prerequisites about the definition of a graph
on an image, the geodesic distances and the exploitation of
redundancy between geodesic propagations, we introduce several
methods of computation of all pairs of distances and we compare
them.

\section{Prerequisites}
\label{sec:pre}

An image $f$ is a discrete function defined on the finite domain $E
\subset \mathbb{N}^2$, with $\mathbb{N}$ the set of positive
integers. The values of a gray level image belongs to $\mathcal{T}
\subset \mathbb{R}$. For a color image (i.e. with 3 channels) the
values are in $\mathcal{T}^3 = \mathcal{T} \times \mathcal{T} \times
\mathcal{T}$ and for a multivariate image of $L$ channels the values
are in $\mathcal{T}^L$. In what follows, we consider $\mathcal{T}
\subset \mathbb{R}^{+}$. The whole results presented in the current
paper are directly extendable to color or multivariate images (Noyel
et. al. \citeyear{Noyel_IAS_2007}, \citeyear{Noyel_ISMM_2007}),
\citep{Noyel_these_2008}.

An image is represented on a grid on which the neighborhood
relations can be defined. Therefore, an image is seen as a non
oriented graph $G=\{V_G,E_G\}$ in which the vertices $V_G$
correspond to the coordinates of pixels, $V_G \in \mathbb{Z} \times
\mathbb{Z}$, and the edges, $E_G \in \mathbb{Z}^2$, give the
neighborhood relations between the pixels. Then, the notion of
neighborhood of a pixel $p$ in the grid is introduced as the set of
pixels which are directly connected to it:

\begin{equation}\label{eq:pre:_voisinage_1}
    \forall p,q \in V_G, \hspace{5mm} p \text{ and } q \text{ are neighbors }
    \Leftrightarrow (p,q) \in E_G,
\end{equation}

\noindent where the ordered pair $(p,q)$ is the edge which joins the
points $p$ and $q$. We assume that a pixel is not its own neighbor
and the neighboring relations are symmetrical. The neighborhood of
pixel $p$, $N_G(p)$, defined a subset of $V_G$ of any size, such as:

\begin{equation}\label{eq:pre:_voisinage_2}
    \forall p,q \in V_G, \hspace{5mm} N_G(p) = \{ q \in V_G, \hspace{1mm} (p,q) \in
    E_G\}.
\end{equation}

Usually in image processing, the following neighborhoods are
defined: 4-neighborhood, 8-neighborhood or 6-neighborhood. In the
sequel, we use the 8-neighborhood. For our study, the choice of the
neighborhood has no influence since we compare several methods using
for each one the same neighborhood.


A path between two points $x$ and $y$  is a chain of points
$\left(x_0, x_1, \ldots, x_i, \ldots, x_l \right) \in E$ such as
$x_0 = x$ and $x_l=y$, and for all $i$, $(x_i, x_{i+1})$ are
neighbours. Therefore, a path $(x_0,x_1, \ldots, x_l)$ can be seen
as a subgraph in which the nodes corresponds to the points and the
edges are the connections between neighbouring points.

The geodesic distance $d_{geo}(x_0,x_l)$, or geodesic time, between
two points of a graph, $x_0$ and $x_l$, is defined as the minimum
distance, or time, between these two points, $\inf_{\mathcal{P}}\{
t_{\mathcal{P}}(x_0,x_l)\}$. The geodesic path $\mathcal{P}_{geo}$
is one of the paths linking these two points with the minimum
distance, $\mathcal{P}_{geo}(x_0,x_l)=(x_0, \ldots, x_l)$:

\begin{eqnarray}\label{eq:pre:_chemin_geo}
  \mathcal{P}_{geo}(x_0,x_l) &=& (x_0, \ldots, x_l) \\
  \text{such as} \hspace{1em}
    d_{geo}(x_0,x_l) &=& \inf_{\mathcal{P}}\{
    t_{\mathcal{P}}(x_0,x_l)\} \nonumber
\end{eqnarray}

If the edges of the graph are weighted by the distance between the
nodes, $t(x_i,x_{i'})$, the geodesic path is one of the sequences of
nodes with the minimum weight.

To generate a geodesic distance, several measures of dissimilarities
can be considered between two neighbors pixels of position $x_i$ and
$x_{i'}$ and of positive grey values $f(x_i)$ and $f(x_{i'})$:
\begin{itemize}
  \item Pseudo-metric $L1$:
  \begin{equation}\label{eq:pre:_distance_abs_difference}
    d_{L_1}( x_i , x_{i'} ) = | f(x_i) - f(x_{i'}) |
\end{equation}
  \item Pseudo-metric sum of grey levels:
  \begin{equation}\label{eq:pre:_distance_somme}
    d_{+}( x_i , x_{i'} ) = f(x_i) + f(x_{i'})
  \end{equation}
\item Pseudo-metric mean of grey levels (similar to the previous pseudo-metric):
  \begin{equation}\label{eq:pre:_distance_moy}
    d_{\overline{+}}( x_i , x_{i'} ) = \frac{f(x_i) + f(x_{i'})}{2}
  \end{equation}
\end{itemize}

The corresponding distances along a path $\mathcal{P} = (x_0,
\ldots, x_l)$ are the sum of the pseudo-metric along this path
$\mathcal{P}$:
\begin{equation}\label{eq:pre:_temps}
    t_{\mathcal{P}}( x_0 , x_{l} ) = \sum_{i=1}^l d( x_{i-1} , x_{i} )
\end{equation}
The geodesic distance is defined as the distance along the geodesic
path. It is also the minimum of distances over all paths connecting
two points $x_0$ and $x_l$:
\begin{eqnarray}\label{eq:pre:_distance}
  d_{geo}( x_0 , x_{l} ) &=& \sum_{i=1}^l\{ d( x_{i-1} , x_{i} ) | x_{i-1} , x_{i} \in \mathcal{P}_{geo}\} \nonumber \\
  {} &=& \inf_{\mathcal{P}}\{t_{\mathcal{P}}( x_0 , x_{l} )\}
\end{eqnarray}

In this paper, we only use the pseudo-metric sum of grey levels
$d_{+}$ which presents the advantage (compared to $d_{L_1}$) not to
be null when two pixels have the same strictly positive value (if
$f(x_i) = f(x_{i'})
> 0$ then $ d_{+}( x_i , x_{i'} ) = 2 f(x_i) > 0 $). The distance
associated to the pseudo-metric sum of grey levels is defined as:
\begin{eqnarray}\label{eq:pre:_distance_geo_somme}
  d_{geo}^{+}( x_0 , x_{l} ) &=& \sum_{i=1}^l d_{+}( x_{i-1} , x_{i} ) \nonumber \\
  {} &=& \sum_{i=1}^l f(x_{i-1}) + f(x_{i})\\
  {} &=& f(x_0) + f(x_l) +  2 \sum_{i=1}^{l-1} f(x_i) \nonumber
\end{eqnarray}

In order to reduce the number of geodesic propagations,
\citet{Bertelli_2006} have used the following properties.

\begin{property}\label{prop:pre:distance_geo}
    Given a geodesic path $(p_0, p_1, \ldots, p_n)$, the geodesic distance,
    $d_{geo}(p_i, p_j)$, between two points $p_i$ et $p_j$ along the path,
    $i < j$, is equal to the difference $d_{geo}(p_0,p_j) -
    d_{geo}(p_0,p_i)$.
\end{property}

In figure \ref{fig:pre:chemin_geo}, if $C$ and $D$ belongs to a
geodesic path connecting $A$ and $B$ the geodesic distance between
$C$ and $D$ is equal to:
\begin{equation}\label{eq:pre:chasles}
\begin{array}{ccccc}
  d_{geo}(C,D) & = & d_{geo}(A,D) &-& d_{geo}(A,C) \\
  8           & = & 20           &-& 12 \\
\end{array}
\end{equation} with $d_{geo}(C,D) = 8$, $d_{geo}(A,D)=20$ and $d_{geo}(A,C)=12$.

\columnbreak

\begin{figure}[!htb]
    \begin{center}
    \includegraphics[width=0.7\columnwidth]{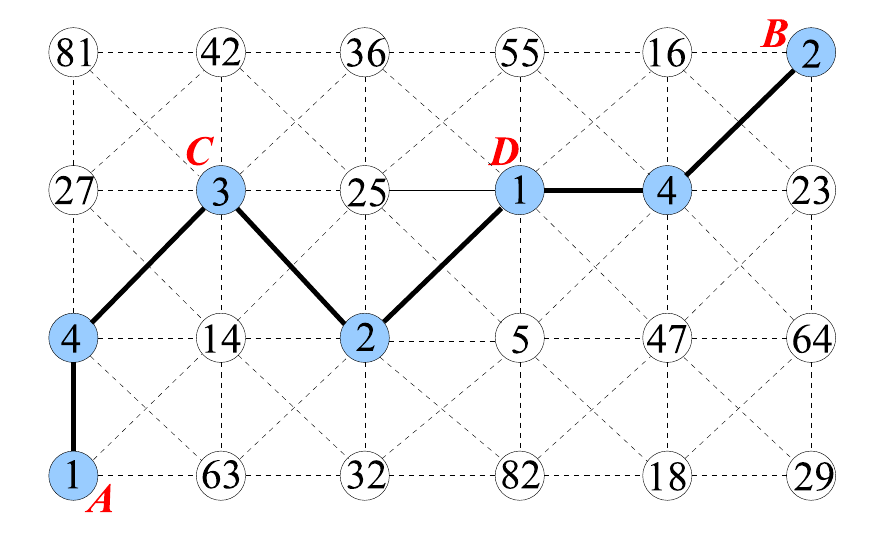}
    \end{center}
    \caption{Discrete geodesic path on a 8-neighborhood graph. The points $C$ and $D$ belong to a geodesic path between
    $A$ et $B$. Therefore the geodesic distance between $C$ and $D$ is also known.}
    \label{fig:pre:chemin_geo}
\end{figure}

Using the property \ref{prop:pre:distance_geo}, when the geodesic
distance between two points of the image is computed, the distances
between all pairs of points along the associated geodesic path are
known. Consequently, the distances array is filled faster using this
redundancy.

In order to compute the distances between the points along the
geodesic path, \citet{Bertelli_2006} proposed to build a geodesic
tree which has three kinds of nodes:
\begin{enumerate}
  \item the root, which is the source point for a geodesic
  propagation. Its distance is null and it has no parents ;
  \item the nodes, which are points having both parents and children
  ;
  \item the leaves, which are points without children.
\end{enumerate}

Starting from the leaves to the nodes (or the opposite), the
distances between points belonging to the same geodesic path are
easily computed.

The following additional property is very useful to compute all
pairs of geodesic distances.

\begin{property}\label{prop:pre:path_length}
    The longer the geodesic paths are, the higher numbers of pairs of
    geodesic distances are computed.
\end{property}

In order to have the benefit of the property
\ref{prop:pre:path_length}, \citet{Bertelli_2006} have chosen as
sources, of the geodesic propagations, random points in a
spiral-like order: starting from the edges of the image and going to
the center. Indeed, the points on the edges of the image tends to
have longer geodesic paths than points located at the center.
Consequently, we want to test their remark by comparing their method
to some others.

In order to make this comparison, we measure the filling rates of
the distances array $D$. For an image containing $N$ pixels, the
distances array is a square matrix of size $N \times N = N^2$
elements. By symmetry, the number of geodesic paths to compute is
equal to:
\begin{equation}\label{eq:pre:_A}
    A = \frac{N^2 - N}{2} 
\end{equation}


The number of computed paths $a$ is determined by counting the
unfilled elements of the distances array $D$. In practice, it is
useful to use a boolean matrix $D_{mrk}$, of size $N \times N$, with
elements equal to 1 if the distance between the pixel located by the
line number and the pixel located by the column number is computed,
and 0 otherwise. By convention in the algorithm, we impose to each
element of the diagonal of the boolean matrix to be equal to one,
$\forall i$ $D_{mrk}(i,i)=1$, because the distance from one pixel to
itself is equal to zero. Due to the symmetric properties of array
$D$, the number of computed paths is equal to:
\begin{eqnarray}\label{eq:pre:_a}
  a &=& \frac{1}{2} \left[ \left(\sum_{k=1}^N \sum_{l=1}^N
D_{mrk}(k,l)\right) - trace(D_{mrk}) \right] \nonumber \\
  {} &=& \frac{1}{2} \left[ \left(\sum_{k=1}^N \sum_{l=1}^N
D_{mrk}(k,l)\right) - N \right]
\end{eqnarray}

Consequently, the filling rate of the distances array is defined by:
\begin{equation}\label{eq:pre:_filling_rate}
    \tau = \frac{a}{A} 
\end{equation}
The proportion of paths to compute for a given pixel $x_i$, is named
the filling rate of the point $x_i$, and is defined by:
\begin{eqnarray}\label{eq:pre:_filling_rate_pt}
  \tau(x_i) &=& \frac{\sum_{k=1}^N D_{mrk}(k,i) - D_{mrk}(i,i)}
    {N-1} \nonumber \\
  {} &=& \frac{\sum_{k=1}^N D_{mrk}(k,i) - 1}
    {N-1}
\end{eqnarray}
When all the distances from one point to the others are computed,
this point is said to be ``filled'', i.e. $\tau(x_i) = 1$.

\columnbreak

\section{Introduction of new methods for fast computation of all pairs of geodesic distances}
\label{sec:method}

In the current section, we initially present \cite{Bertelli_2006}
method, named the ``spiral method'', and then we introduce two new
methods before making comparisons : 1) a geodesic extrema method and
2) a method based on the filling rate of all distances pairs array.
The empirical comparisons are made on three different images of size
$25 \times 25$ pixels: ``bumps'', ``hairpin bend'', ``random'' (fig.
\ref{fig:method:_im_test}).

\begin{figure}[!htb]
    \centering
  \begin{tabular}{@{}c@{ }c@{ }c@{}}
    \includegraphics[width=0.29\columnwidth]{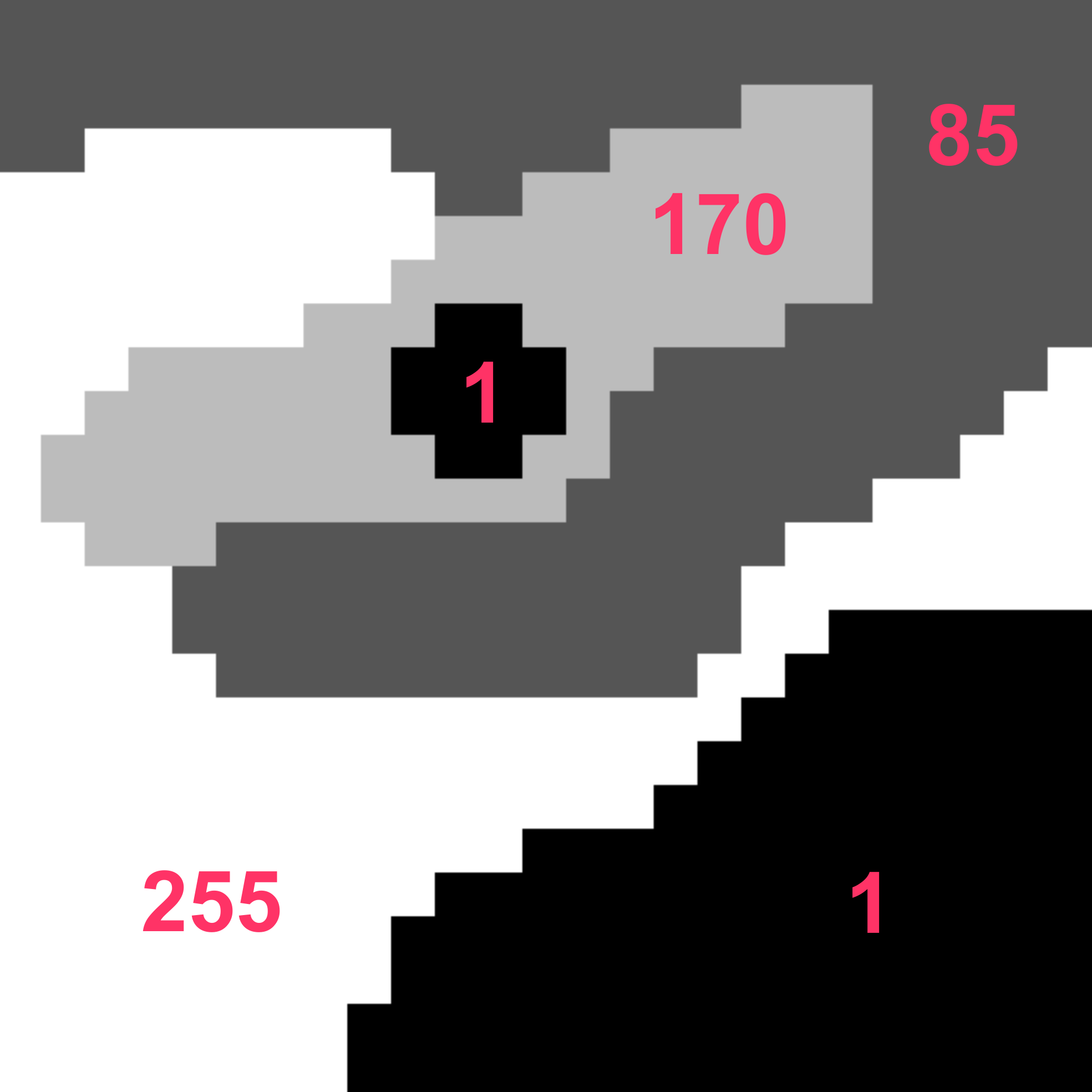}&
    \includegraphics[width=0.29\columnwidth]{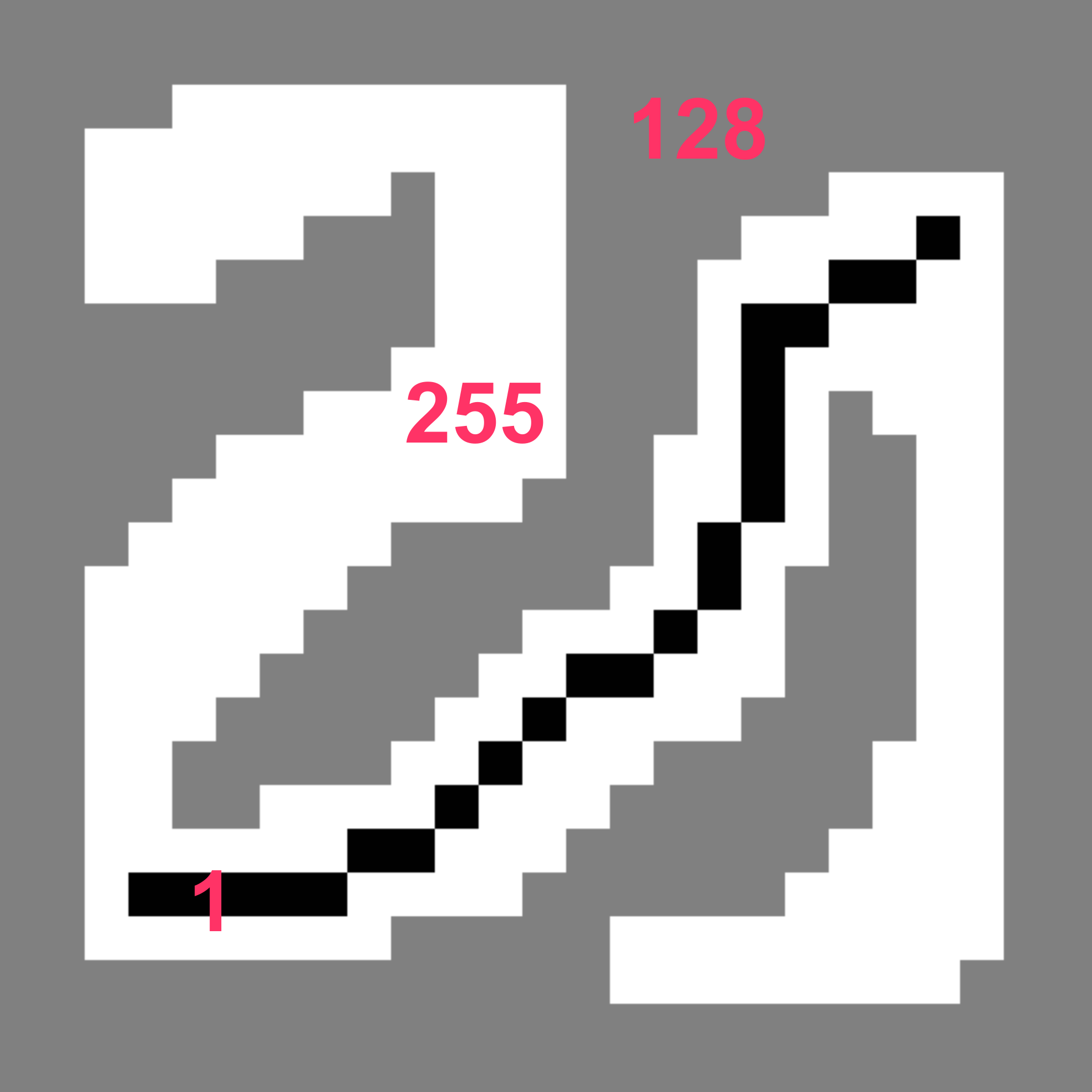}&
    \includegraphics[width=0.3\columnwidth]{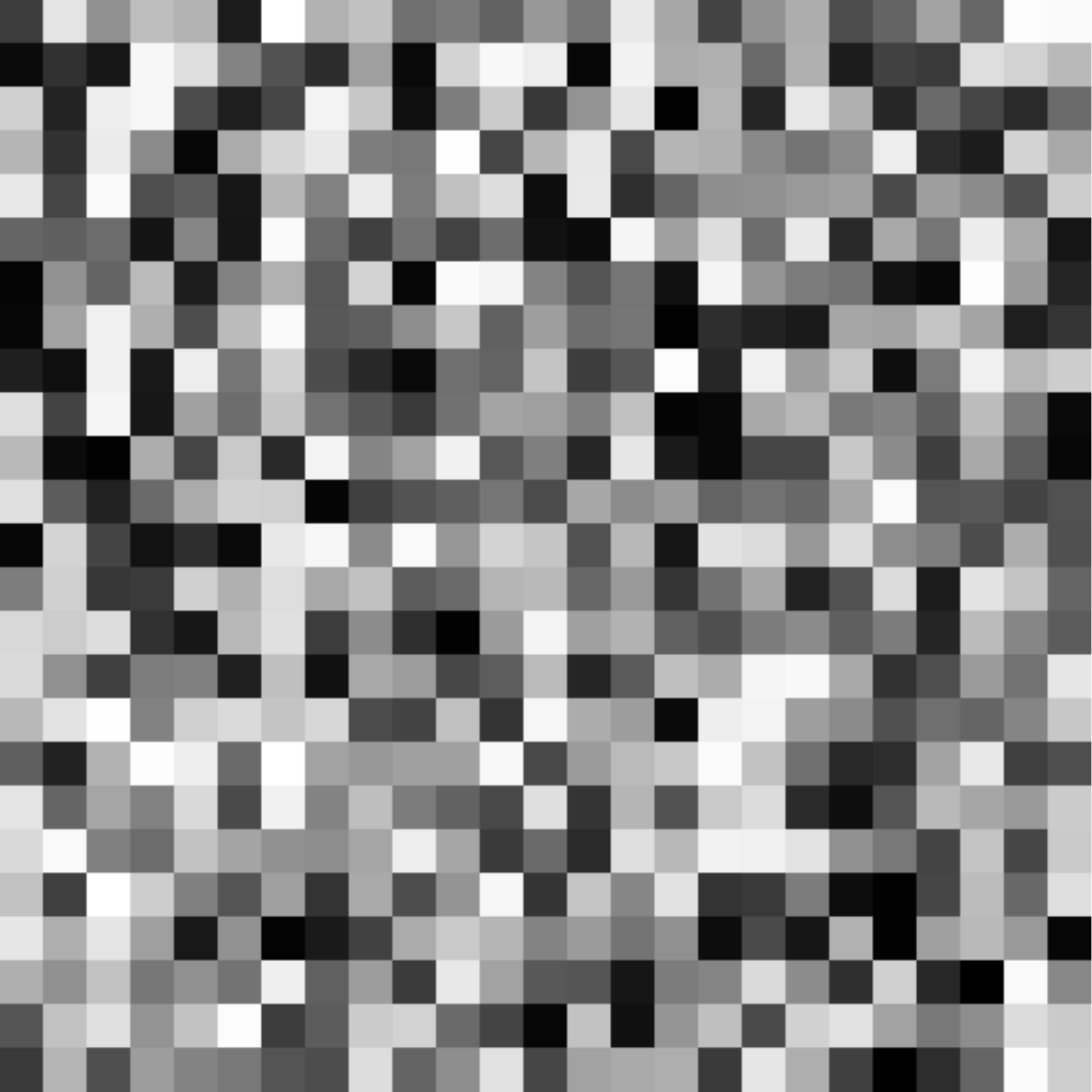}\\
    \footnotesize Image ``bumps'' & \footnotesize Image ``hairpin bend'' & \footnotesize Image ``random''\\
  \end{tabular}
  \caption{Images of size $25 \times 25$ pixels ``bumps'', ``hairpin bend'' and
  ``random'' whose grey levels are between 1 and 255.}
  \label{fig:method:_im_test}
\end{figure}

In order to use homogeneous measures for all methods, the Soille's
algorithm \citep{Soille_2003}, called ``geodesic time function'' is
used, with a discrete neighborhood of size $3 \times 3$ pixels. In
fact, we do not need an Euclidean geodesic algorithm to make this
comparison study. The Euclidean version is described in
\citep{Soille_1992} and improved in \citep{Coeurjolly_2004}.

\subsection{Spiral method}

\citet{Bertelli_2006} affirms that the source points of the geodesic
propagations, with the longest paths, are in general situated on the
borders of the image. As these points are useful to reduce the
number of operations, the source points are chosen in a random way
on concentric spiral turns of image pixels. A concentric spiral turn
is a frame, of one pixel width, in which the top left corner is at
position (1,1) or (2,2) or (3,3) or etc. The figure
\ref{fig:method:_turns} gives an example. While not all the pixels
of the spiral turns have been selected, we draw one pixel, among
them, in a uniform random way ; otherwise we switch to the next
spiral turn.

\begin{figure}[!htb]
    \centering
    \includegraphics[width=0.3\columnwidth]{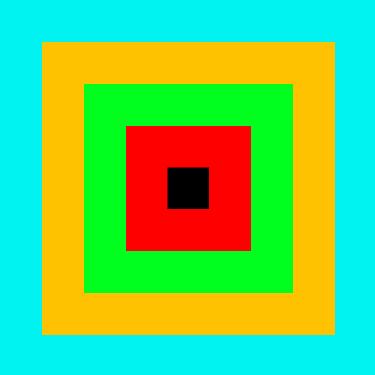}
  \caption{The spiral turns of an image $9 \times 9$ pixels.}
  \label{fig:method:_turns}
\end{figure}

\columnbreak

\begin{table}[!htb]
\begin{tabular}{p{\columnwidth}}
\hline \caption{Algorithm: Spiral method}\label{alg:method:_spiral}\\
\hline
\end{tabular}
\begin{algorithmic}[1]
    \STATE Given $D$ the distances array of size $N \times N$
    \WHILE{ $D$ is not filled}
        \STATE Select the most exterior spiral turn not yet filled
        \STATE Determine $S$ the list of pixels of the spiral turn not yet
        filled
        \WHILE{ $S$ is not empty }
            \STATE Select a source point $s$ randomly in $S$
            \STATE Compute the geodesic tree from $s$
            \STATE Fill the distances array $D$
            \STATE Remove the points of $S$ which are filled
        \ENDWHILE
    \ENDWHILE
\end{algorithmic}

\begin{tabular}{p{\columnwidth}}
\\
\hline
\end{tabular}
\end{table}


For each image, the filling rate is plotted versus the number of
propagations (fig. \ref{fig:method:_filling_rate_spiral}). The
number of propagations which are necessary to fill the distances
array by the spiral method and the naive method are also given in
this figure. The relative difference of the number of propagations
of the spiral method compared to the number of propagations of the
naive method is written $\Delta_r(naive)$. We notice that the spiral
method reduces the number of propagations by a factor ranging
between 13.4 \% and 25.6 \%, as compared to the naive one.
Consequently, it is very useful to exploit the redundancy in the
propagations by building a geodesic tree.


\begin{figure}[!htb]
    \centering
  \begin{tabular}{cc}
    \includegraphics[width=0.4\columnwidth]{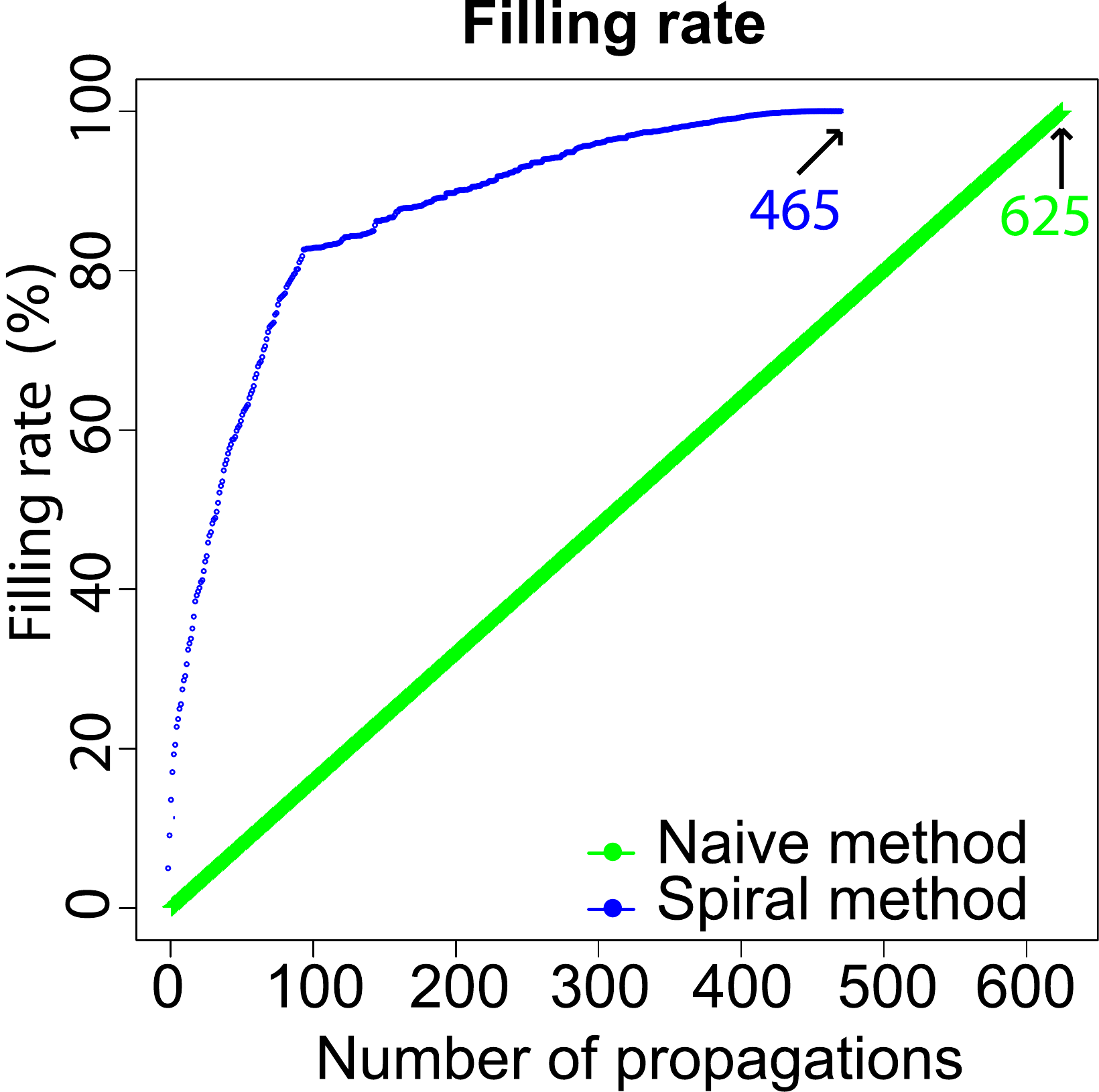}&
    \includegraphics[width=0.4\columnwidth]{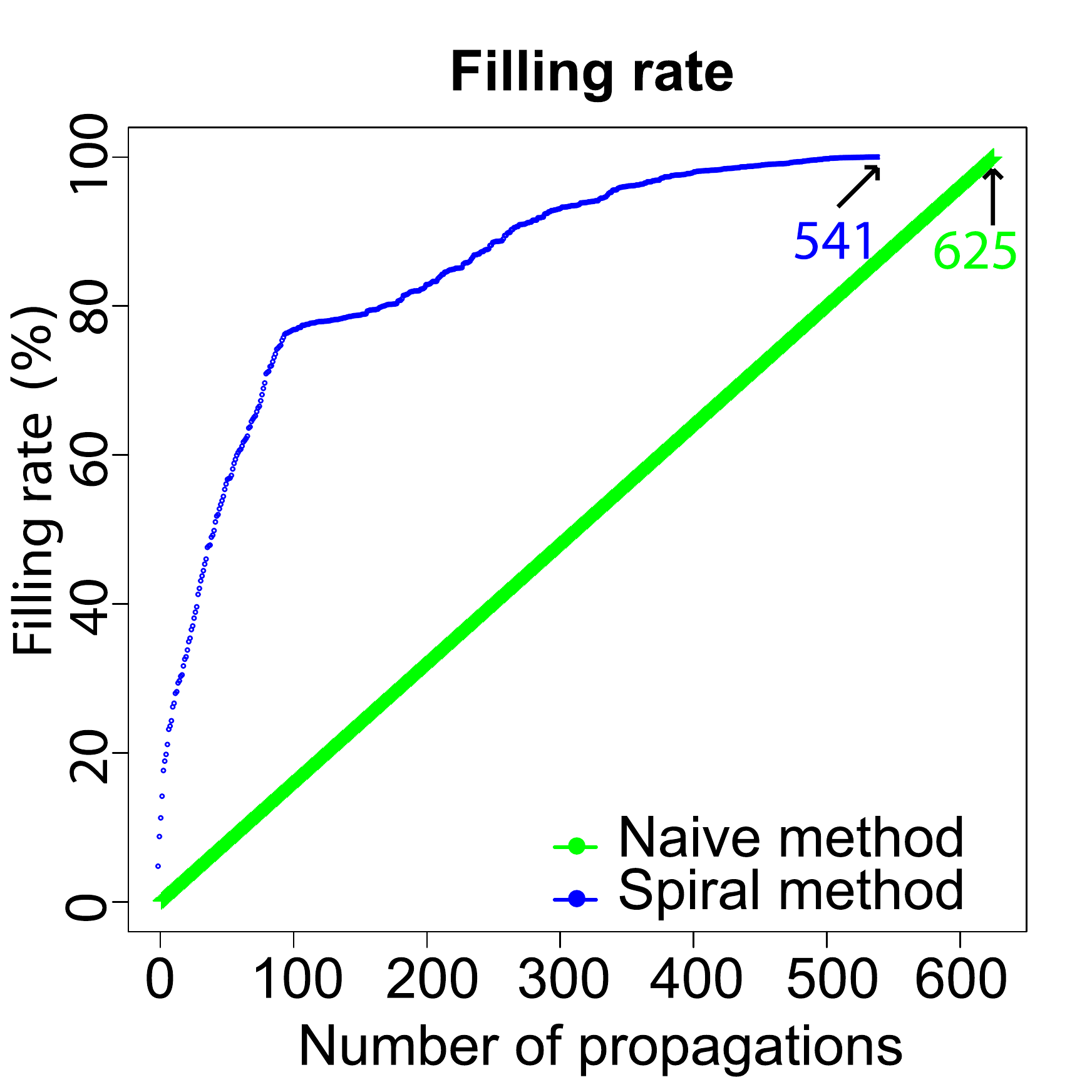}\\
    \footnotesize (a) ``bumps'' & \footnotesize (b) ``hairpin bend'' \\
    \footnotesize $\Delta_r(naive) = 25.6 \%$ &
    \footnotesize $\Delta_r(naive) = 13.4 \%$ \\
    \includegraphics[width=0.4\columnwidth]{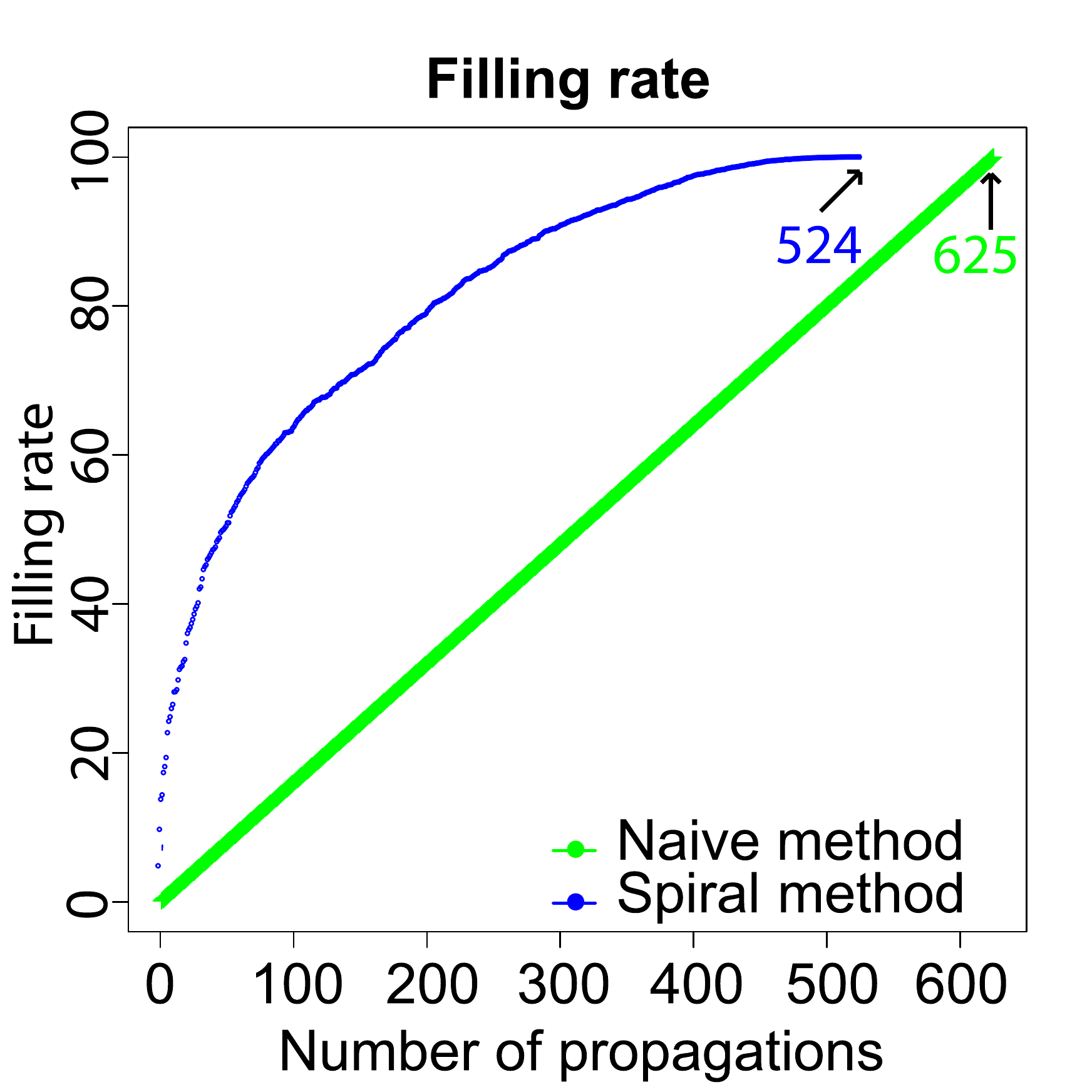}\\
    \footnotesize (c) ``random''\\
    \footnotesize $\Delta_r(naive) = 16.2 \%$ \\
  \end{tabular}
  \caption{Comparison of the filling rates $\tau$ of the distances array,
  between the naive method (in green) and the spiral method (in blue)
  for the images ``bumps'', ``hairpin bend'' and ``random''.
  The relative differences $\Delta_r(naive)$ of the number of
  propagations of the spiral method compared to the number of propagations of the
  naive method are given on the bottom line.}
  \label{fig:method:_filling_rate_spiral}
\end{figure}

\FloatBarrier

\subsection{Spiral method with repulsion}

Two neighbours points have a high probability to have similar
geodesic trees. Consequently, a first improvement of the spiral
method is to introduced a repulsion distance between the points
drawn randomly in a spiral like-order (algorithm of table
\ref{alg:method:spiral_repulsion}). Several tests have shown us that
a repulsion distance of three pixels on both sides of a source point
gives the best filling rates.

These tests are empirical tests. In fact, several repulsion
distances were tried and it has been noticed that the value of three
pixels gives the best results in order to fill the array of
distances. This value of three pixels is certainly related to the
image size, because, generally, the farther the source points of the
geodesic propagations are the faster the array of all pairs of
geodesic distances is filled.

\begin{table}[!htb]

\begin{tabular}{p{\columnwidth}}
\hline \caption{Algorithm: Spiral method with repulsion}\label{alg:method:spiral_repulsion}\\
\hline
\end{tabular}

\begin{algorithmic}[1]
    \STATE Given $D$ the distances array of size $N \times N$
    \STATE Given $h$ the repulsion distance: $h$ $\leftarrow$ 3
    pixels
    \WHILE{ $D$ is not filled }
        \STATE Select the most exterior spiral turn not yet filled
        \STATE Determine $S$ the list of pixels of the spiral turn not yet
        filled
        \WHILE{ $S$ is not empty }
            \STATE Select a source point $s$ randomly in $S$
            \STATE Remove in the list $S$ the two left points of $s$ and the
            two right points of $s$ if they are still in $S$
            \STATE Compute the geodesic tree from $s$
            \STATE Fill the array $D$
            \STATE Remove the points of $S$ which are filled
        \ENDWHILE
    \ENDWHILE
\end{algorithmic}
\begin{tabular}{p{\columnwidth}}
\\
\hline
\end{tabular}
\end{table}

The figure \ref{fig:method:_spiral_repulsion}, shows that the
relative differences in the number of propagations necessary to fill
the distances array are larger than 15 \% for images ``bumps'' and
``hairpin bend'' and than 5.3 \% for the image ``random''. Therefore,
the spiral method with the repulsion distance is faster than the
spiral method to fill the distances array.

\columnbreak

\begin{figure}[!htb]
    \centering
  \begin{tabular}{cc}
    \includegraphics[width=0.4\columnwidth]{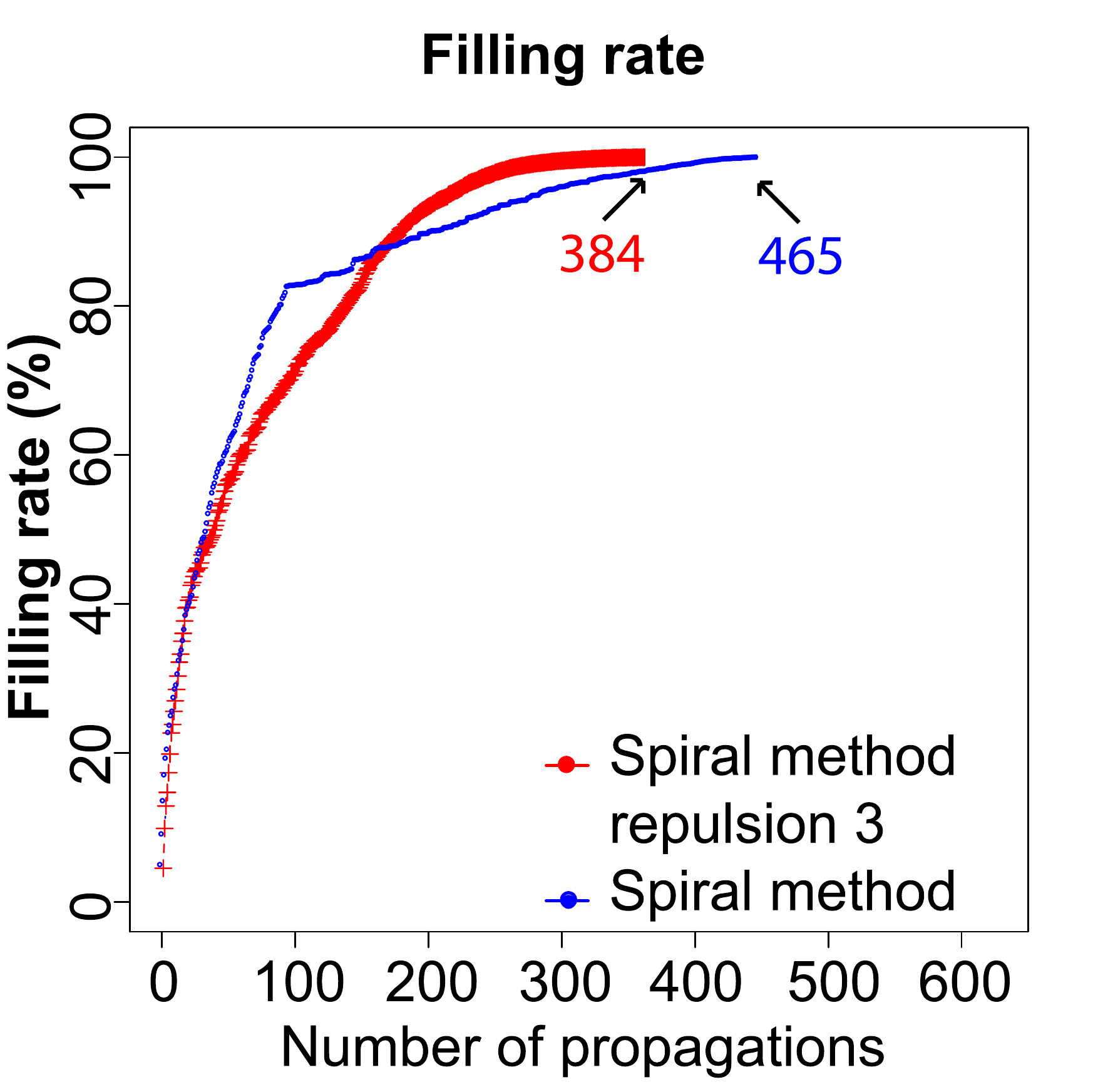}&
    \includegraphics[width=0.4\columnwidth]{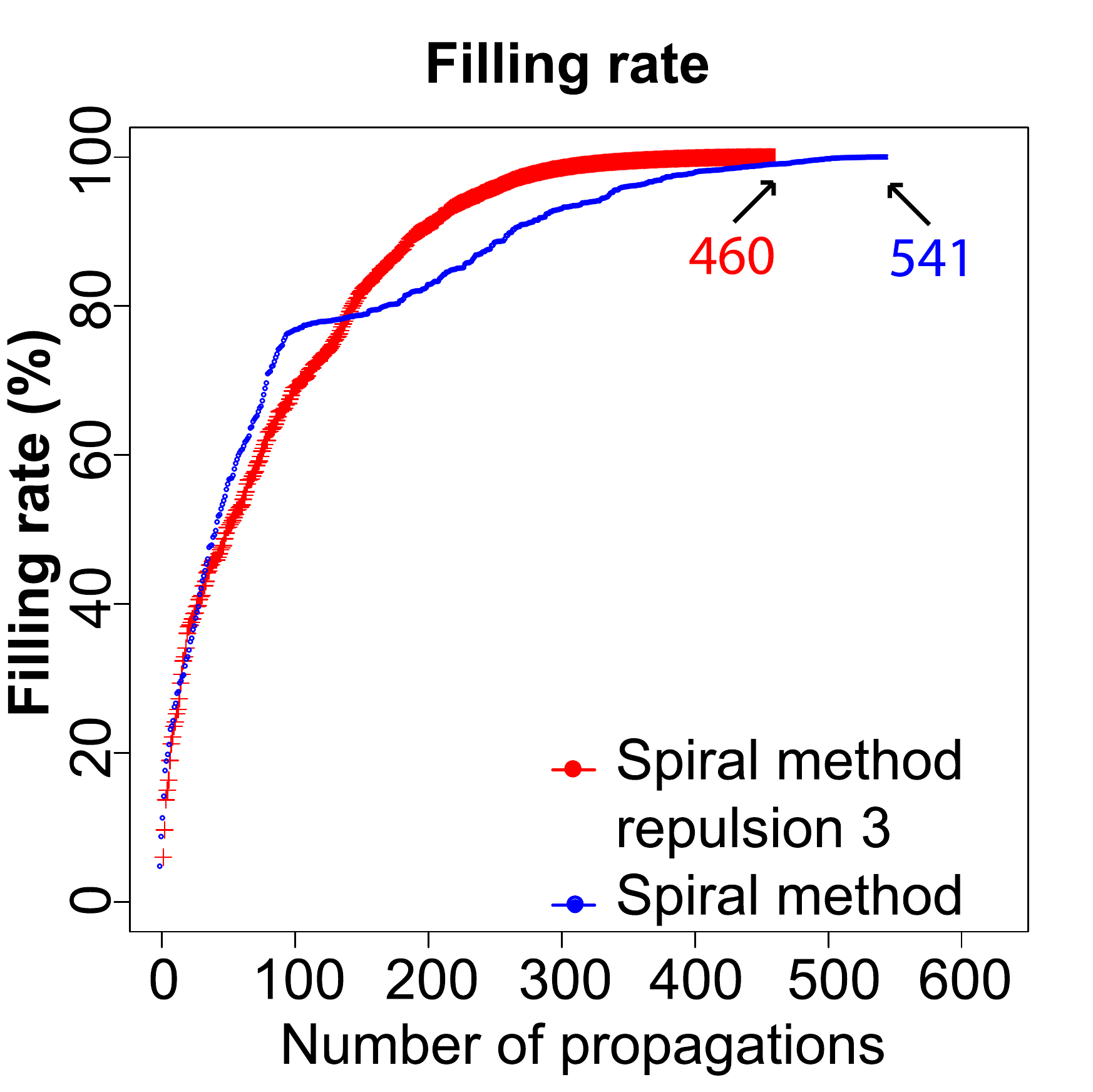}\\
    \footnotesize (a) ``bumps'' & \footnotesize (b) ``hairpin bend''\\
    \footnotesize $\Delta_r(spiral) = 17.4 \%$ &
    \footnotesize $\Delta_r(spiral) = 15.0 \%$ \\
    \footnotesize $\Delta_r(naive)  = 38.6 \%$ &
    \footnotesize $\Delta_r(naive)  = 26.4 \%$ \\
    \includegraphics[width=0.4\columnwidth]{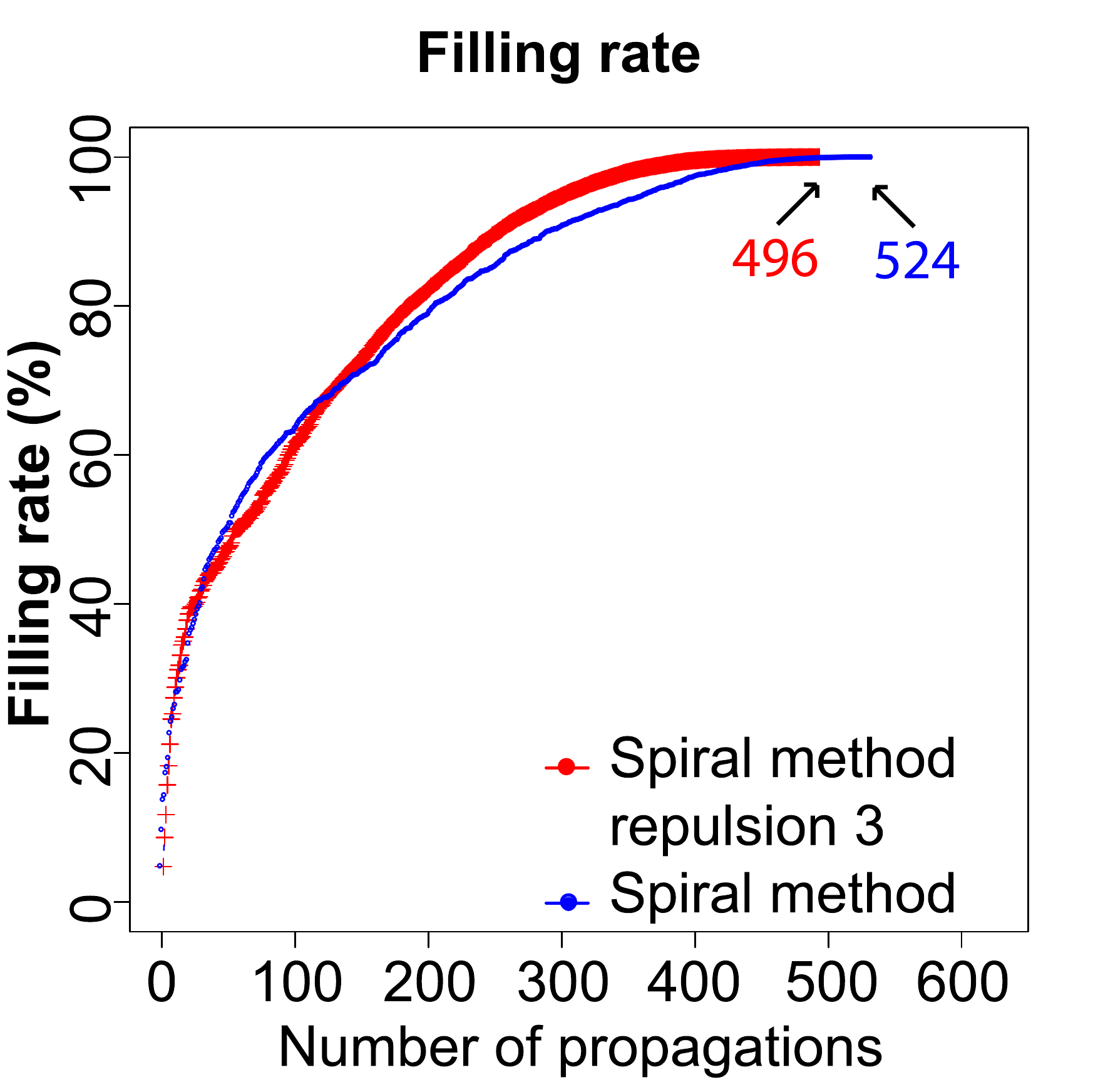}\\
    \footnotesize (c) ``random''\\
    \footnotesize $\Delta_r(spiral) = 5.3  \%$ \\
    \footnotesize $\Delta_r(naive)  = 20.6 \%$ \\
  \end{tabular}
  \caption{Comparison of the filling rates $\tau$ of the distances array,
  between the spiral method with repulsion (in red)
  and the spiral method (in blue)
  for the images ``bumps'', ``hairpin bend'' and ``random''.
  The relative differences $\Delta_r(spiral)$ (resp. $\Delta_r(naive)$) of the number of
  propagations of the spiral method with repulsion compared to the number
  of propagations of the spiral (resp. naive) method are given on the bottom lines.}
  \label{fig:method:_spiral_repulsion}
\end{figure}

\subsection{Geodesic extrema method}

As the longest geodesic paths are those which fill the most the
distance array, we look for the geodesic extrema of the image. To
compute the geodesic extrema of the image, we use two geodesic
propagations:
\begin{itemize}
  \item a first propagation starts from the edges of the
  image. Then we select one of the points $c$ with the longest distance from
  the edges. This point is called the geodesic centroid ;
  \item a second propagation from the geodesic centroid gives
  the farthest points from the centroid, i.e. the geodesic
  extrema of the image.
\end{itemize}
On the figure \ref{fig:method:_extrema_maps}, which shows the
results from these two propagations, we notice that the geodesic
extrema are mainly on the edges of the image, which ascertains the
motivation to use a spiral method.

Then the pixels are sorted by geodesic distance from the centroid
into the list of geodesic extrema $Ext$. This list is used to choose
the source points of the geodesic propagations. If several points
have the same geodesic distance from the centroid, then the less
filled is selected (algorithm of table \ref{alg:method:extrema}).

\begin{figure}[!htb]
    \centering
  \begin{tabular}{@{}c@{ }c@{ }c@{}}
    \includegraphics[width=0.29\columnwidth]{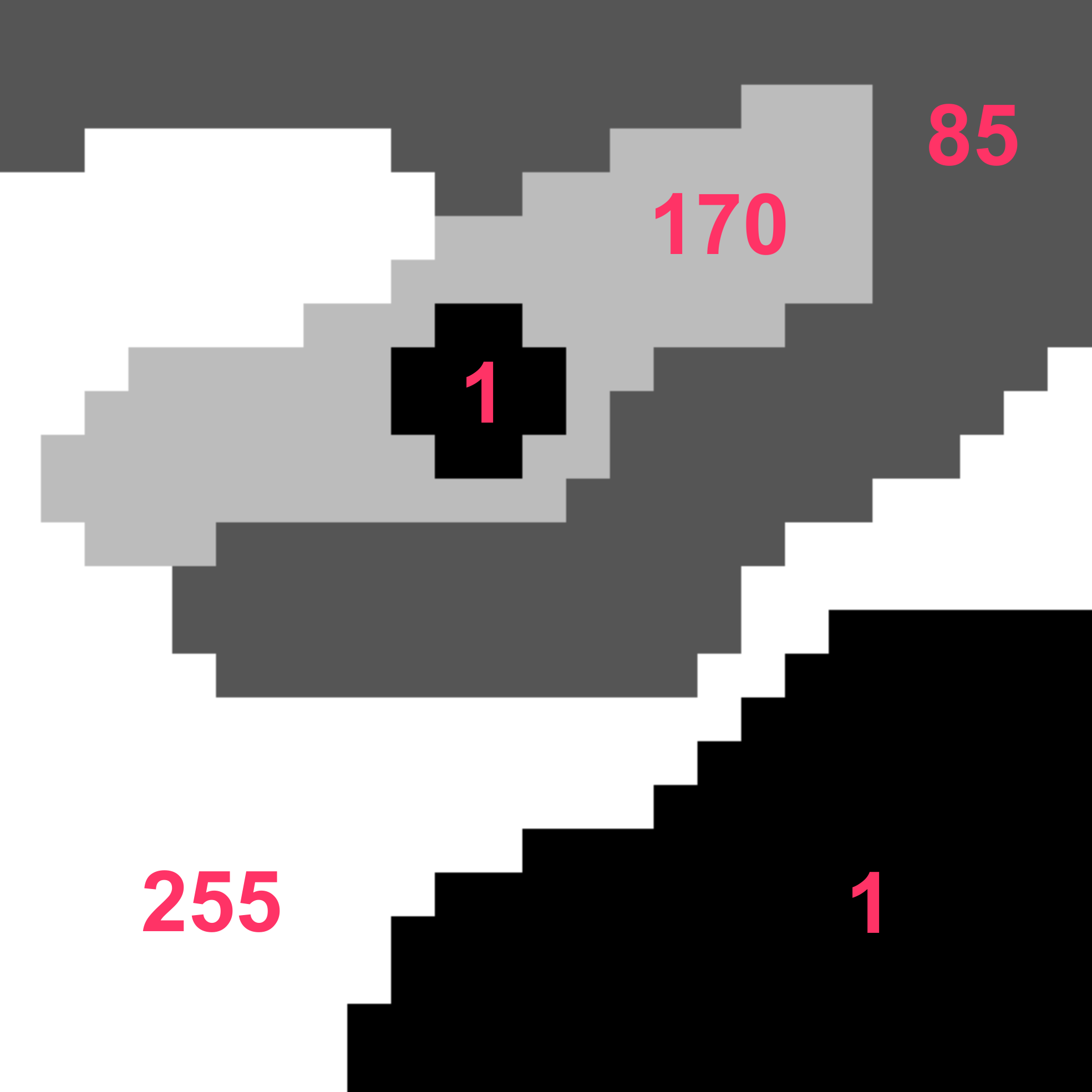}&
    \includegraphics[width=0.29\columnwidth]{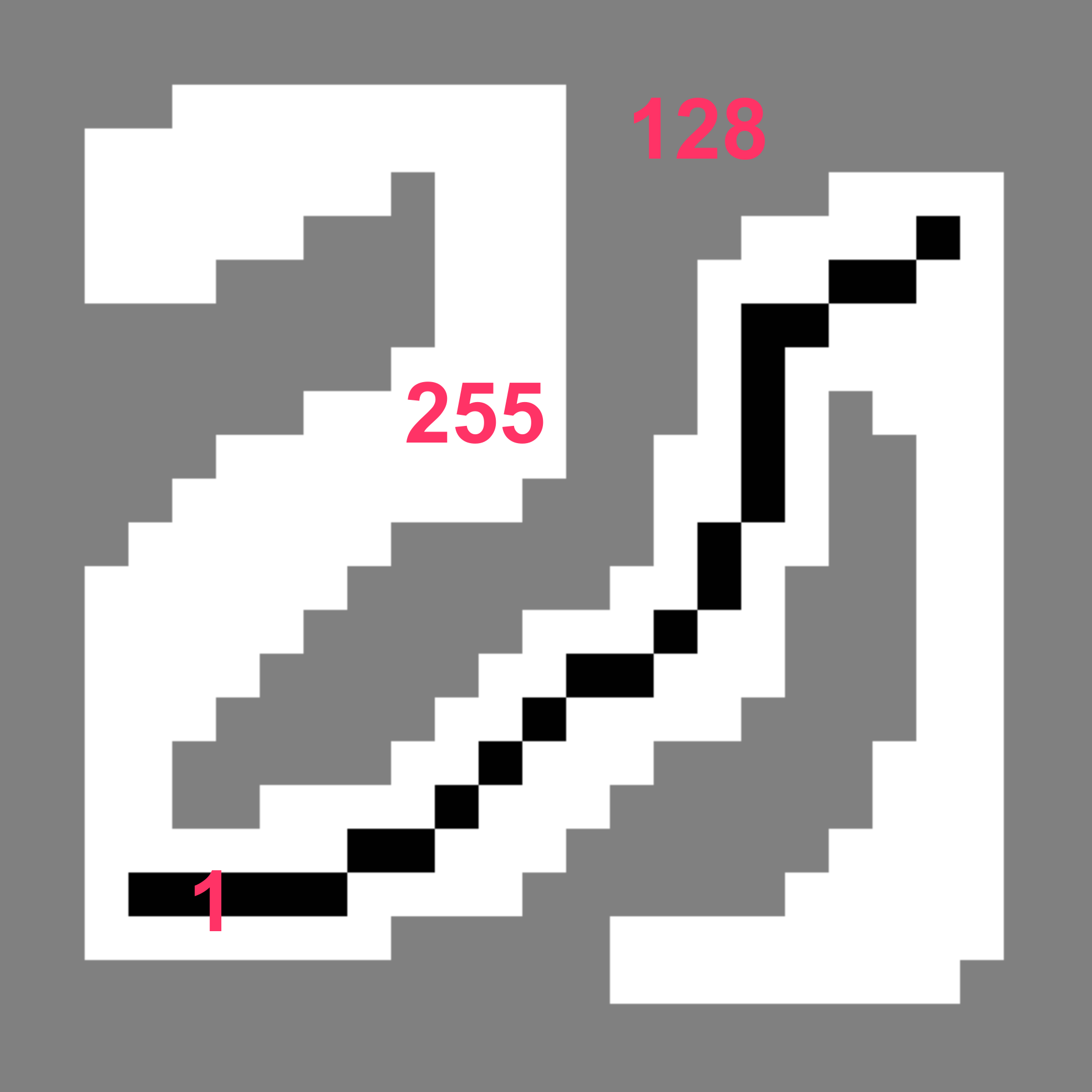}&
    \includegraphics[width=0.3\columnwidth]{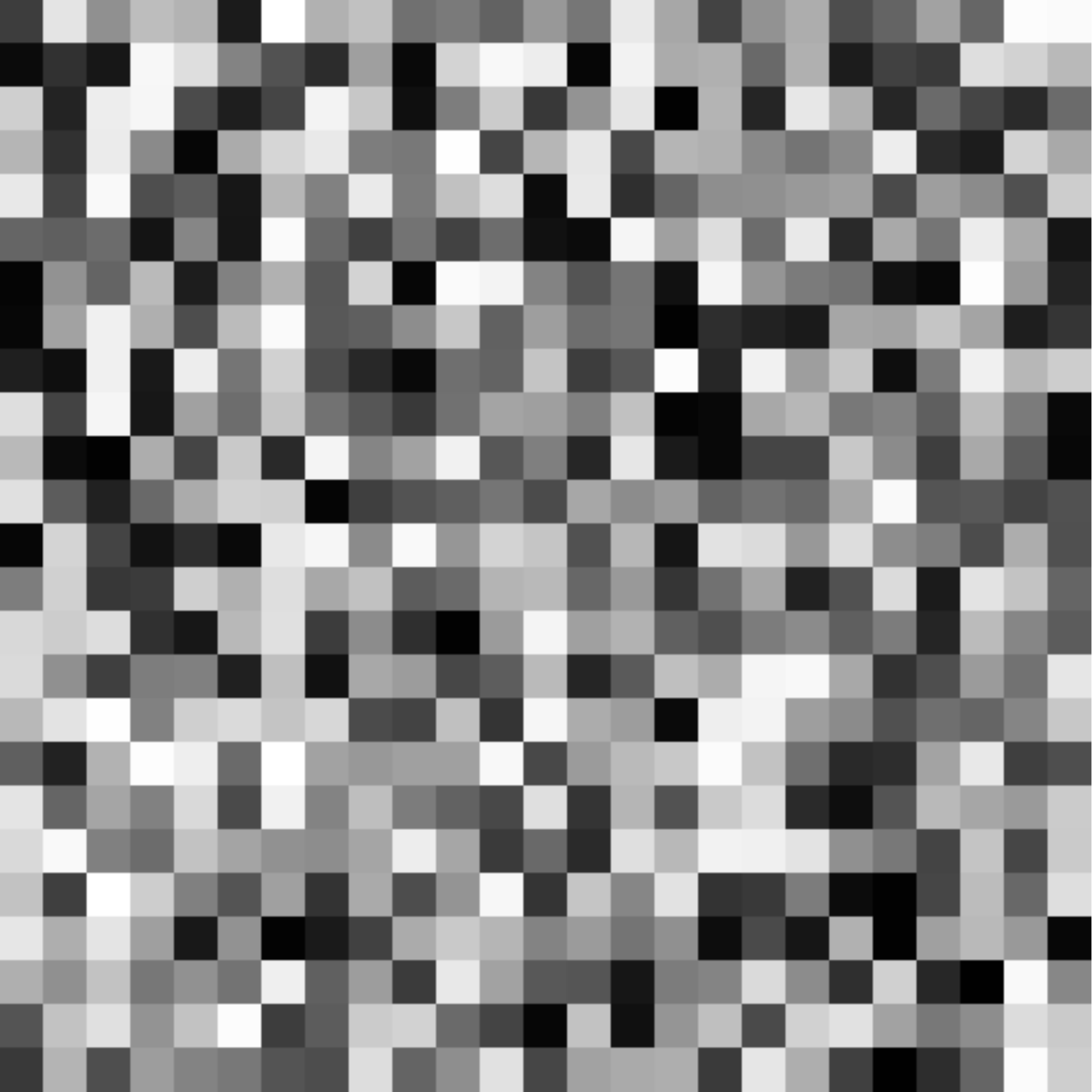}\\
    \footnotesize ``bumps'' & \footnotesize ``hairpin bend'' & \footnotesize ``random''\\
    \includegraphics[width=0.3\columnwidth]{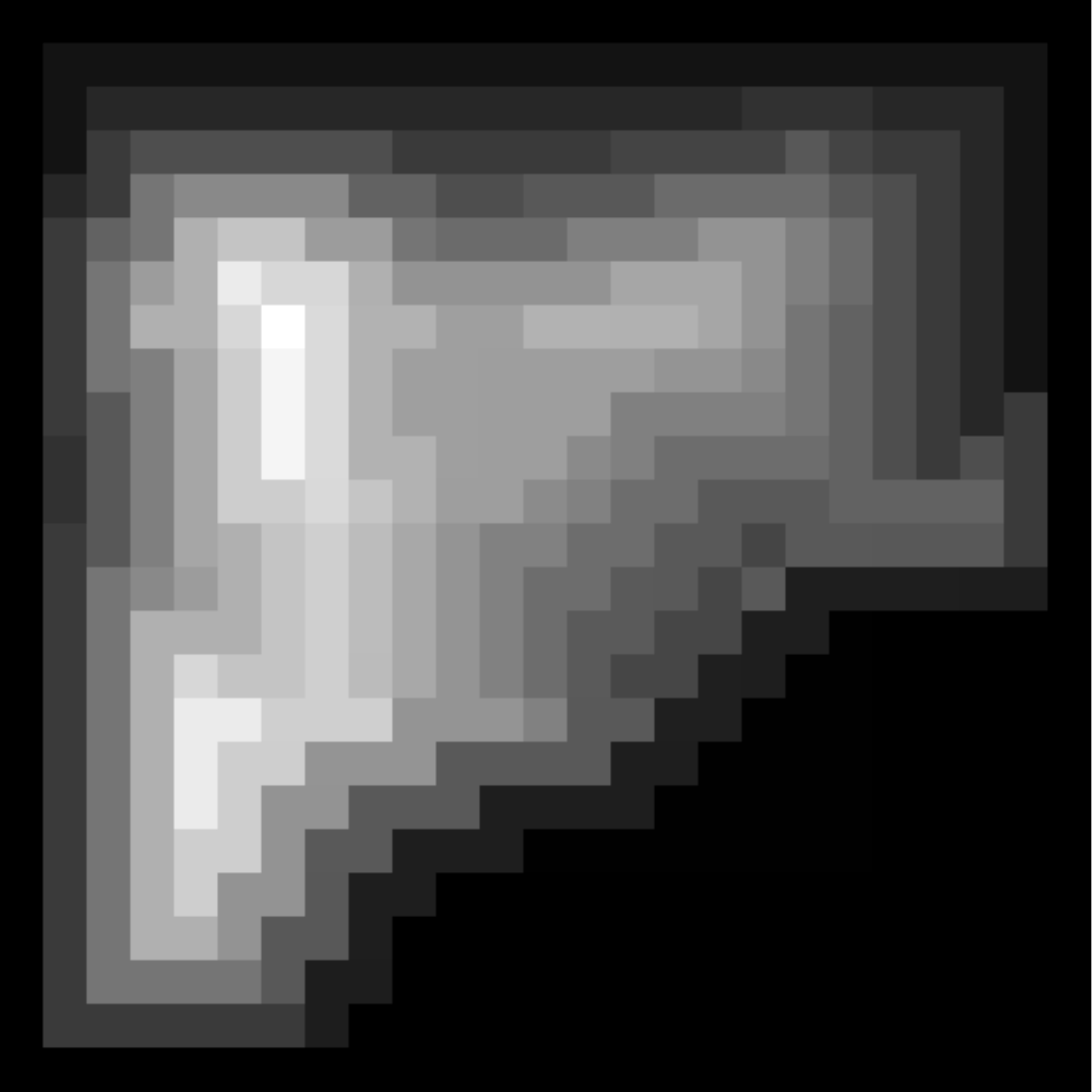}&
    \includegraphics[width=0.3\columnwidth]{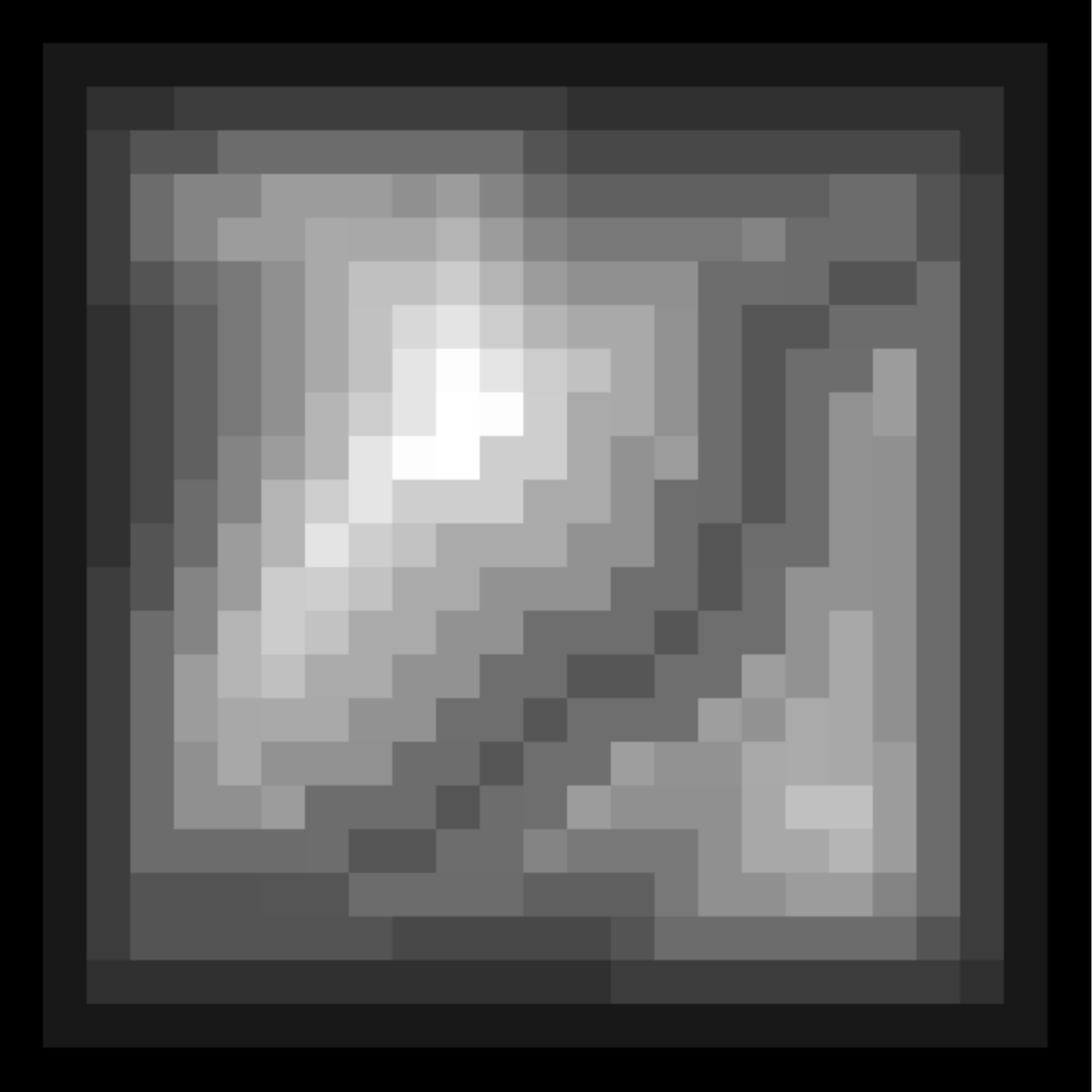}&
    \includegraphics[width=0.3\columnwidth]{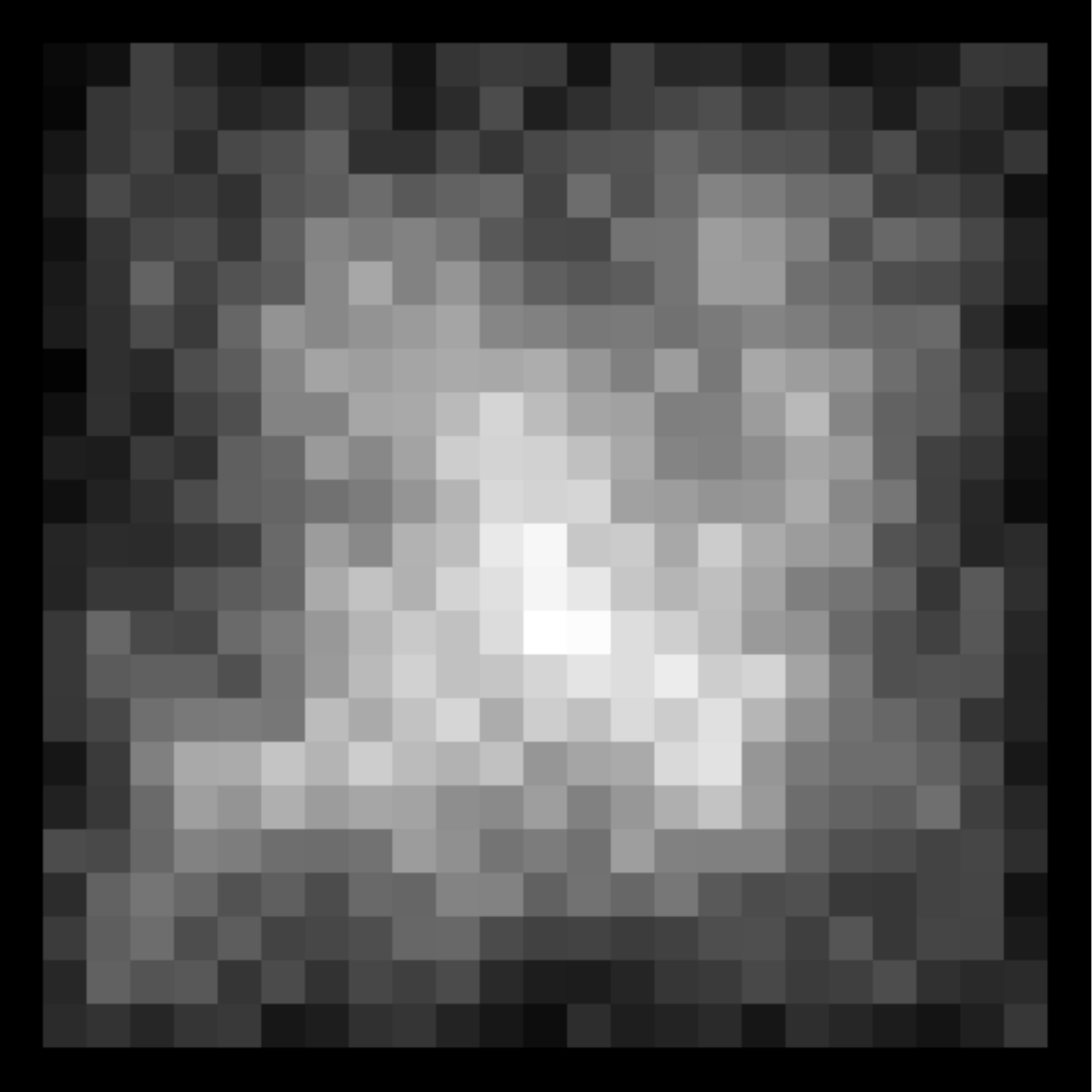}\\
    \multicolumn{3}{c}{\footnotesize Geodesic distance from the edges of the image}\\
    \includegraphics[width=0.3\columnwidth]{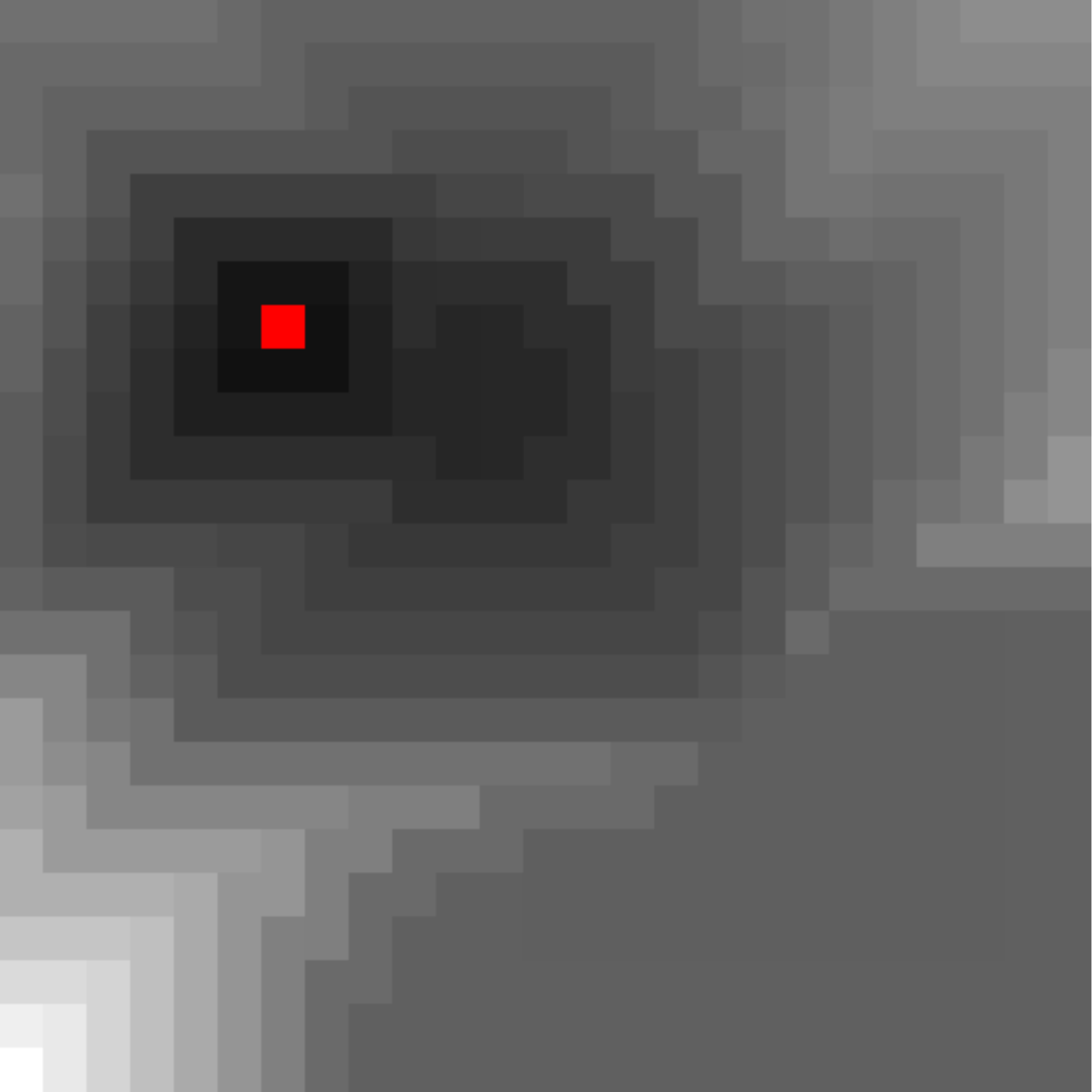}&
    \includegraphics[width=0.3\columnwidth]{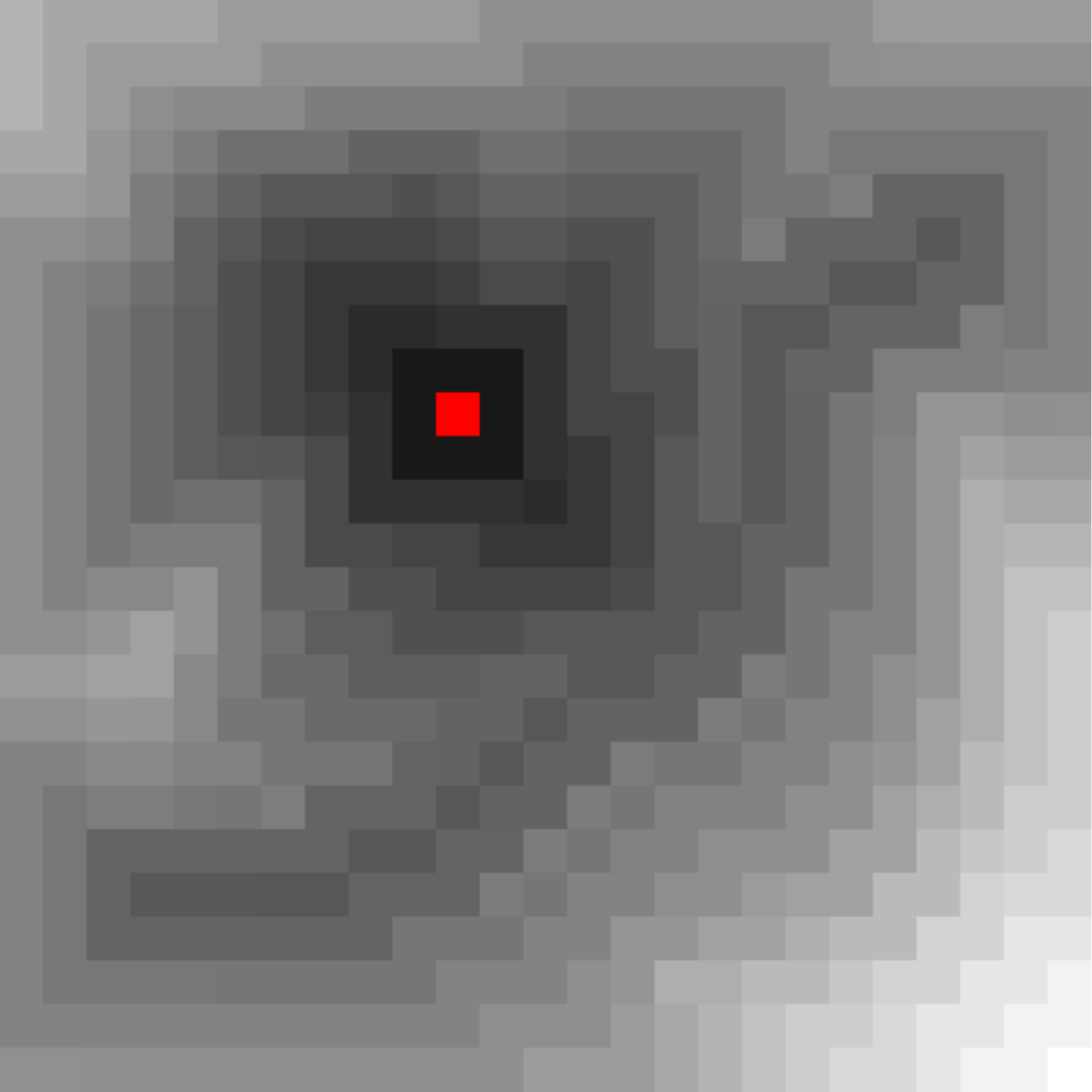}&
    \includegraphics[width=0.3\columnwidth]{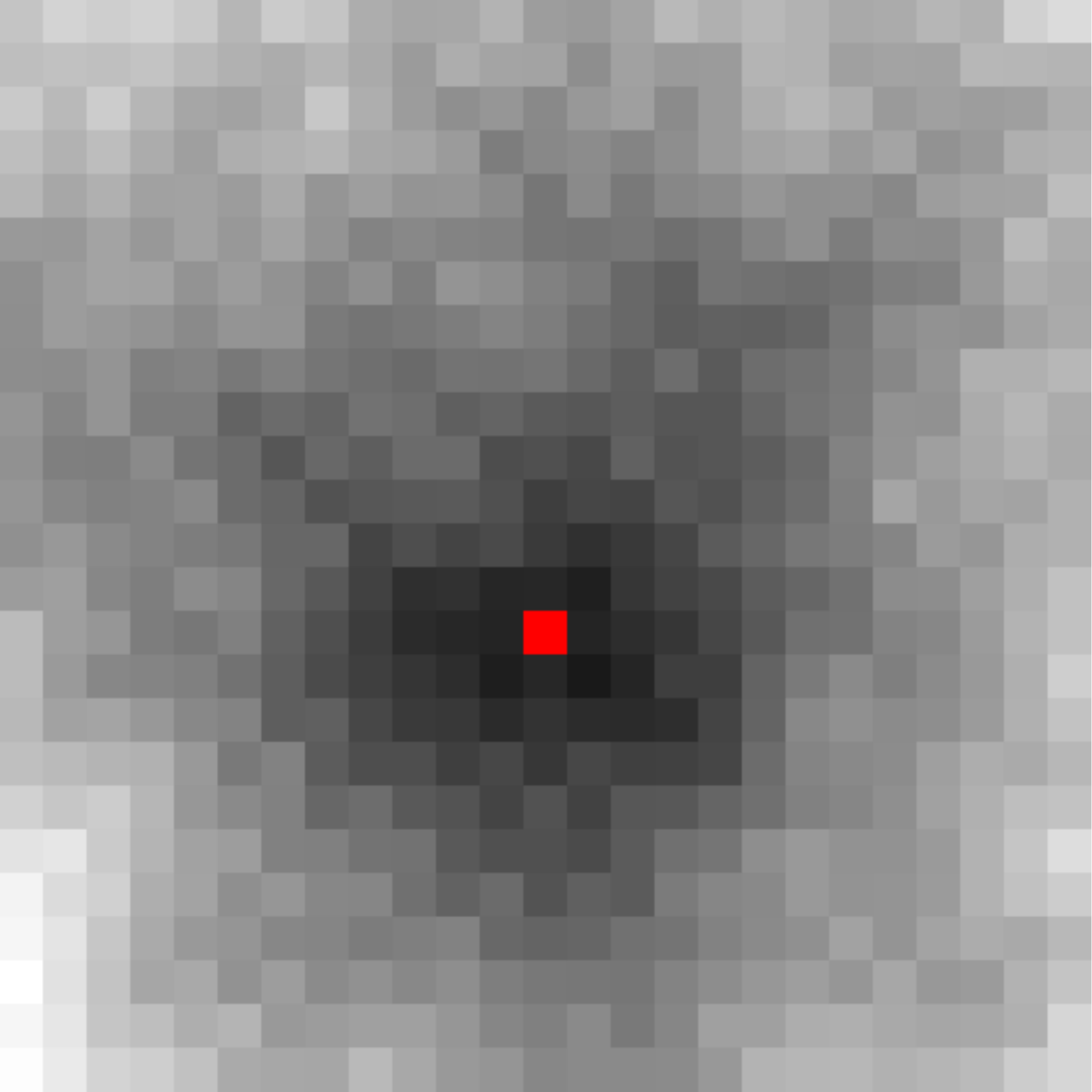}\\
    \multicolumn{3}{c}{\footnotesize Geodesic distance from the centroid (red point)}\\
  \end{tabular}
  \caption{Geodesic distance from the edges of the image (second line)
  and from the centroid in red (third line) for several images.}
  \label{fig:method:_extrema_maps}
\end{figure}

\begin{table}[!htb]
\begin{tabular}{p{\columnwidth}}
\hline \caption{Algorithm: Geodesic extrema method} \label{alg:method:extrema}\\
\hline
\end{tabular}
\algsetup{indent =2em}
\begin{algorithmic}[1]
    \STATE Given $D$ the distances array of size $N \times N$
    \STATE Compute the geodesic centroid $c$
    \STATE Compute the decreasing list of geodesic extrema $Ext$
    whose first element $Ext[1]$ is the greatest geodesic extrema not
    yet filled
    \STATE Initialise to zero, the list, of size $N$, of the filling rate of points
    $\tau$.
    \WHILE{ $D$ is not filled}
        \STATE $B$ $\leftarrow$ $\{x \in Ext | d_{geo}(x,c) = d_{geo}(Ext[1],c) \}$
        \STATE $s$ $\leftarrow$ $x$
        \STATE Compute the geodesic tree from $s$
        \STATE Fill the distances array $D$
        \STATE Remove the points of the list $Ext$ which are filled
        \STATE Update the list of filling rates $\tau$
    \ENDWHILE
\end{algorithmic}
\begin{tabular}{p{\columnwidth}}
\\
\hline
\end{tabular}
\end{table}

\columnbreak

\begin{figure}[!htb]
    \centering
  \begin{tabular}{cc}
    \includegraphics[width=0.4\columnwidth]{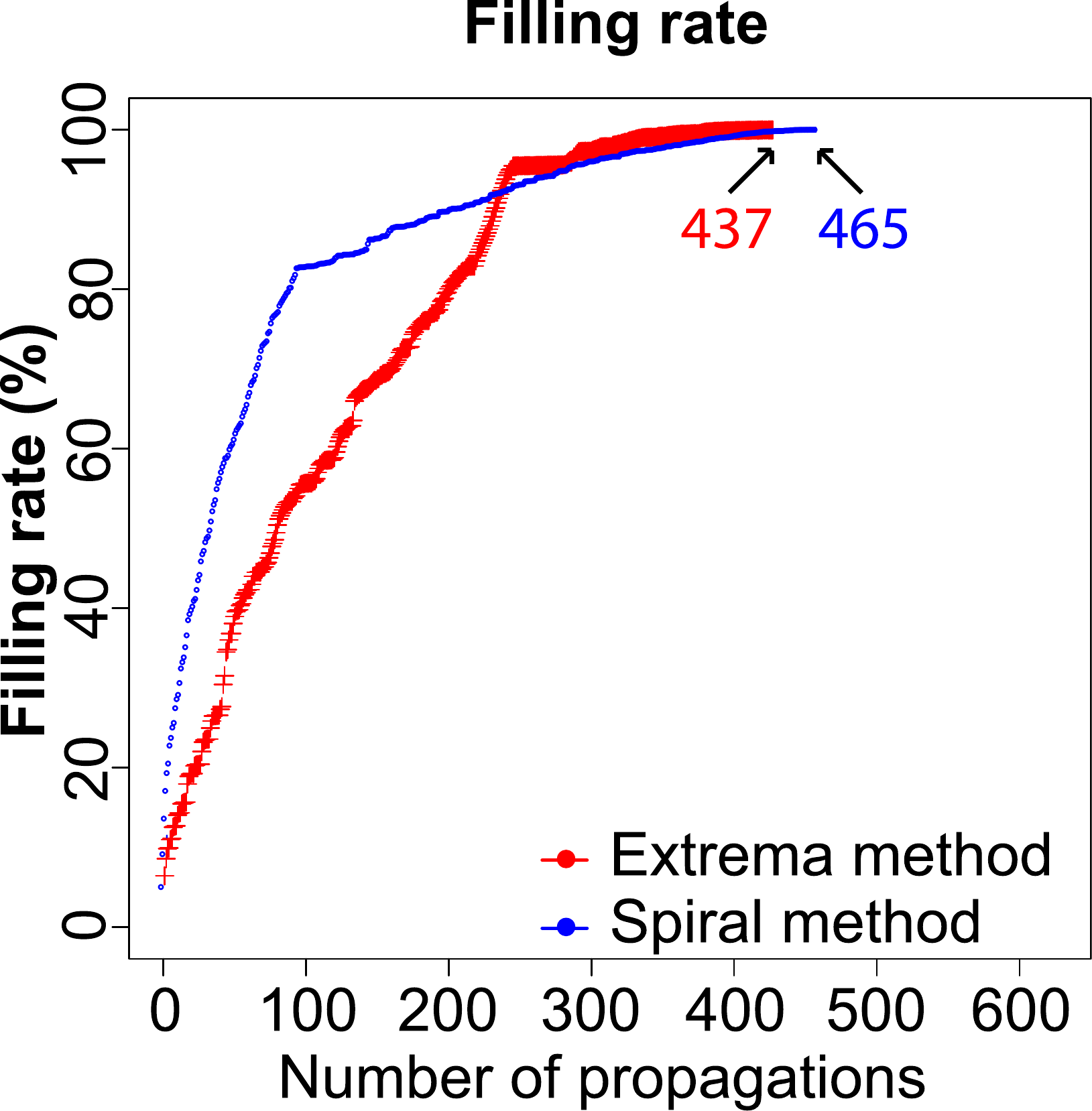}&
    \includegraphics[width=0.4\columnwidth]{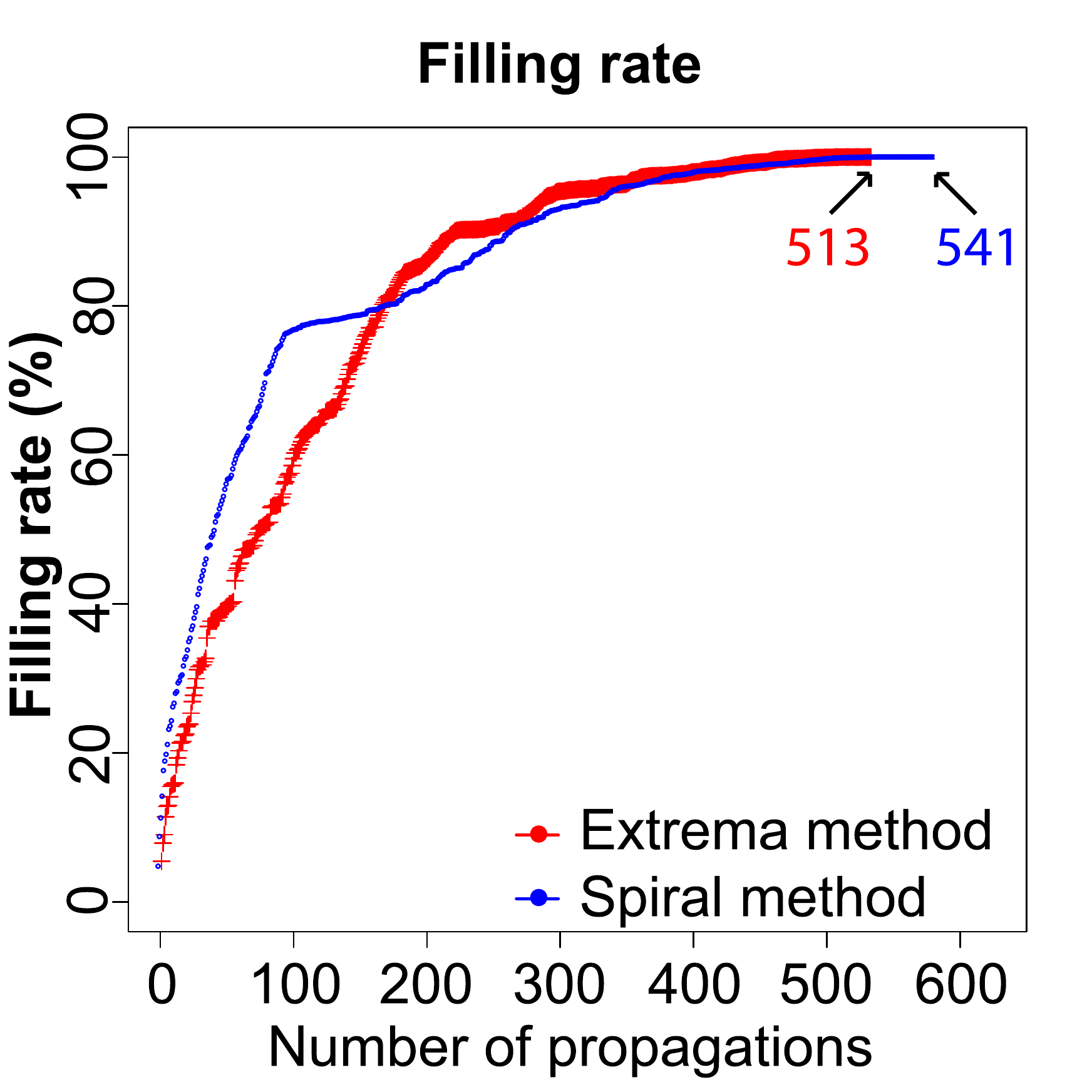}\\
    \footnotesize (a) ``bumps'' & \footnotesize (b) ``hairpin bend''\\
    \footnotesize $\Delta_r(spiral) = 6.0 \%$ &
    \footnotesize $\Delta_r(spiral) = 5.2 \%$ \\
    \footnotesize $\Delta_r(naive)  = 30.1 \%$ &
    \footnotesize $\Delta_r(naive)  = 17.9 \%$\\
    \includegraphics[width=0.4\columnwidth]{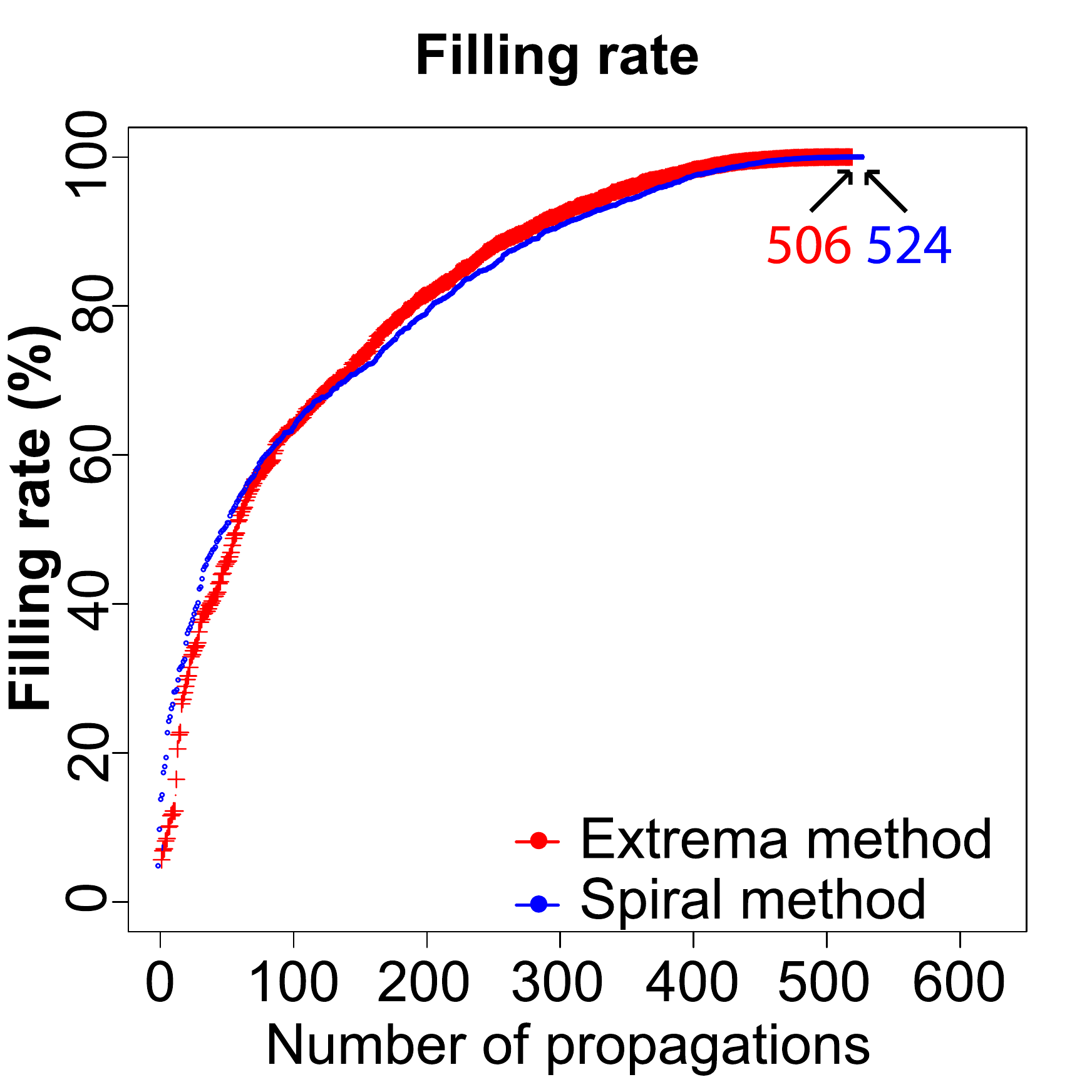}\\
    \footnotesize (c) ``random''\\
    \footnotesize $\Delta_r(spiral) = 3.4 \%$ \\
    \footnotesize $\Delta_r(naive)  = 19.0 \%$ \\
  \end{tabular}
  \caption{Comparison of the filling rates $\tau$ of the distances array,
  between the geodesic extrema method (in red) and the spiral method (in blue)
  for the images ``bumps'', ``hairpin bend'' and ``random''.
  The relative differences $\Delta_r(spiral)$ (resp. $\Delta_r(naive)$) of the number of
  propagations of the the geodesic extrema method compared to the number of propagations of the spiral
  (resp. naive) method are on the bottom lines.}
  \label{fig:method:_filling_rate_extrema}
\end{figure}

The filling rates of the spiral method and the geodesic extrema
method versus the number of propagations are plotted for each image
(fig. \ref{fig:method:_filling_rate_extrema}). We notice that the
filling rates of the geodesic extrema method are at the beginning
inferior or similar to these of the spiral method. However, at the
end the filling rates of the geodesic extrema method are better than
these of the spiral method. In fact, we are looking for an approach
filling totally the distances array in the fastest way. As the
number of propagations of the geodesic extrema method necessary to
fill the distances array are lower than in the spiral method, the
geodesic extrema method fills faster the distances array than the
spiral method.

In order to get an exact comparison it is necessary to generate two
propagations, corresponding to the determination of the geodesic
extrema, to the number of propagations necessary to fill the
distances array. Even, with this modification, the extrema method
fills the distances array faster than the spiral method (the
relative differences $\Delta_r(spiral)$ are between 3.4 \% and 6
\%).

\subsection{Method based on the filling rate of the distances array}

In place of selecting the source points from their distance from the
geodesic centroid propagation, we select first the less filled
points. To this aim, after each geodesic propagation the filling
rate is computed for each point. Then the less filled point is
selected as a source of the propagation. If several points are among
the less filled, then the greatest geodesic extrema is chosen among
these points (algorithm of table \ref{alg:method:filling}).

\begin{table}[!htb]
\begin{tabular}{p{\columnwidth}}
\hline \caption{Algorithm: Method based on the filling rate of the distances array}\label{alg:method:filling}\\
\hline
\end{tabular}

\begin{algorithmic}[1]
    \STATE Given $D$ the distances array of size $N \times N$
    \STATE Compute the decreasing list of geodesic extrema $Ext$
    \STATE Initialise to zero, the list, of size $N$, of the filling rate of points
    $\tau$.
    \WHILE{ $D$ is not filled }
        \STATE $B$ $\leftarrow$ $\{ \argmin_{x \in E} \tau[x] \}$
        \IF{ $Card\{B\} > 1$}
            \STATE $s$ $\leftarrow$  $\argmin_{x \in B} Ext[x]$
        \ELSE
            \STATE $s$ $\leftarrow$ $\argmin_{x \in E} \tau[x]$
        \ENDIF
        \STATE Compute the geodesic tree from $s$
        \STATE Fill the distances array $D$
        \STATE Remove the points of the list $Ext$ which are filled
        \STATE Update the list of filling rates $\tau$
    \ENDWHILE
\end{algorithmic}
\begin{tabular}{p{\columnwidth}}
\\
\hline
\end{tabular}
\end{table}

As for the geodesic extrema method, we compare the filling rates of
the method based on the filling rate of the distances array and the
spiral method versus the number of propagations (fig.
\ref{fig:method:_filling_rate}). We notice that the method based on
the filling rate of $D$ reduces of 32.3 \% the number of
propagations of the spiral method on the image ``bumps'' and 18.7 \%
on the image ``hairpin bend''. Consequently this method fills faster
the distances array than the spiral method. Even for the image
``random'' the method based on the filling rate of $D$ still improves
the spiral method of 2.7 \%. However it is not very common to
compute all pairs of geodesic distances on a strong unstructured
image such as the ``random'' one.

By comparison to the naive approach, the method based on the filling
rate of $D$ reduces the number of propagations by a  49.6 \% rate
(resp. 29.6 \%) on the image ``bumps'' (resp. ``hairpin bend'').

As for the previous method, in order to get an exact comparison, it
is necessary to add two propagations, corresponding to the
determination of the geodesic extrema, to the number of propagations
necessary to fill the distances array. Even, with this modification,
the number of propagations necessary to fill the distances array is
still lower for the method based on the filling rate of $D$.

\begin{figure}[!htb]
    \centering
  \begin{tabular}{cc}
    \includegraphics[width=0.4\columnwidth]{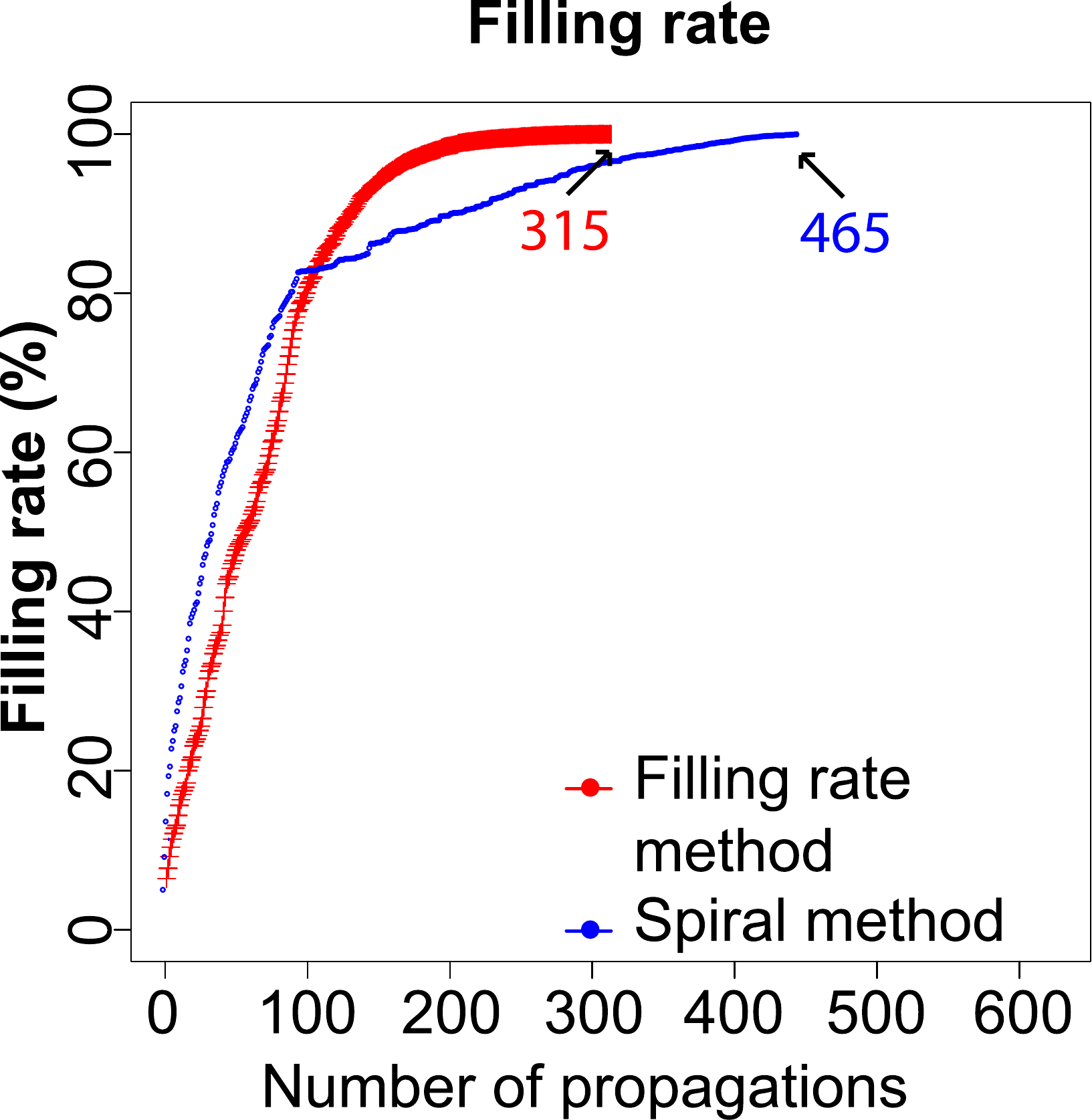}&
    \includegraphics[width=0.4\columnwidth]{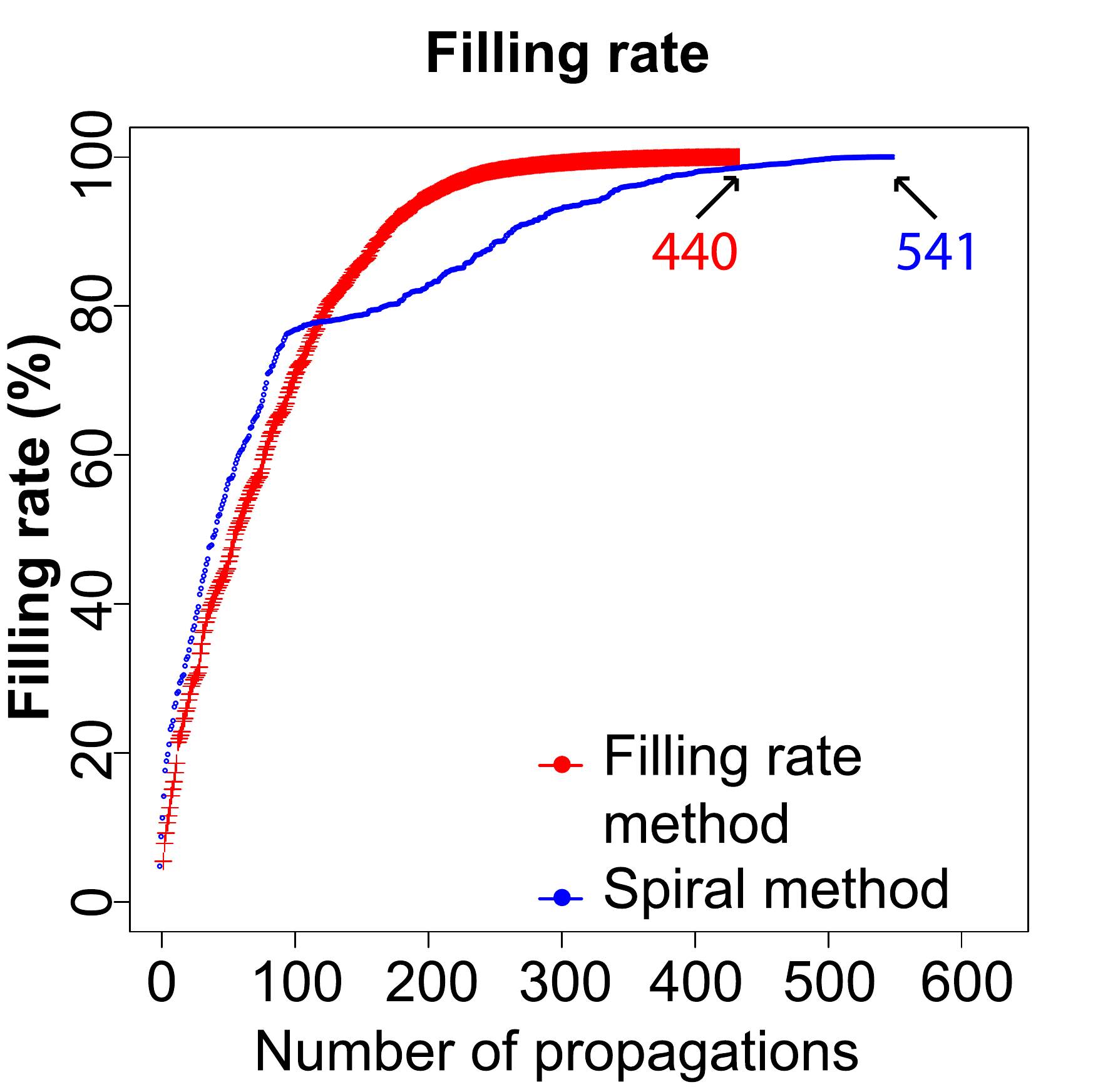}\\
    \footnotesize (a) ``bumps'' & \footnotesize (b) ``hairpin bend''\\
    \footnotesize $\Delta_r(spiral) = 32.3 \%$ &
    \footnotesize $\Delta_r(spiral) = 18.7 \%$\\
    \footnotesize $\Delta_r(naive)  = 49.6 \%$ &
    \footnotesize $\Delta_r(naive)  = 29.6 \%$\\
    \includegraphics[width=0.4\columnwidth]{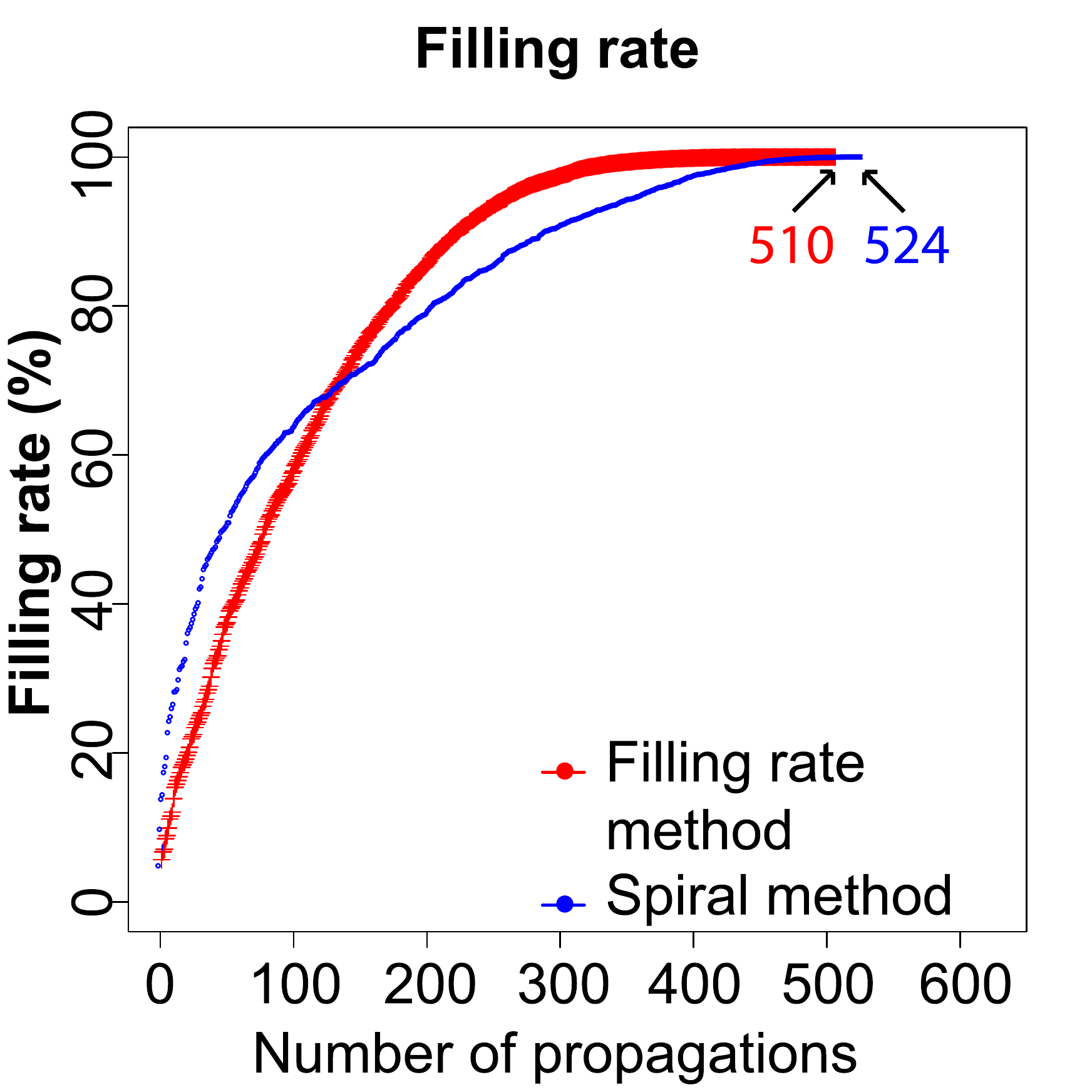}\\
    \footnotesize (c) ``random''\\
    \footnotesize $\Delta_r(spiral) = 2.7  \%$ \\
    \footnotesize $\Delta_r(naive)  = 18.4 \%$ \\
  \end{tabular}
  \caption{Comparison of the filling rates $\tau$ of the distances array,
  between the method based on the filling rate of the distances array (in red)
  and the spiral method (in blue)
  for the images ``bumps'', ``hairpin bend'' and ``random''.
  The relative differences $\Delta_r(spiral)$ (resp. $\Delta_r(naive)$) of the number of
  propagations of the method based on the filling rate compared to the number of propagations
  of the spiral (resp. naive) method are given on the bottom lines.}
  \label{fig:method:_filling_rate}
\end{figure}

\FloatBarrier
\section{Discussion}
\label{sec:comparaison}

After having presented and tested several methods to fill the
distances array, we have compared them for the three test images
``bumps'', ``hairpin bend'' and ``random'' on the figure
\ref{fig:comparaison:comparaison} and in the table
\ref{tab:comparaison:comparaison}. For the images ``bumps'' and
``hairpin bend'', the method based on the filling rate of the
distances array is faster than the other algorithms. For the
``random image'' (an extreme case presenting no texture) the methods
introduced here give similar results, since the relative difference
between the maximum number of propagations is less than 2.7 \%.
Therefore, we conclude that the method based on the filling rate of
the distances array is the best one to calculate the array of all
pairs of distances. According to their performances, the others are
ranked in the following order: 1) the spiral method with repulsion
distance, 2) the geodesic extrema method and 3) the spiral method of
\citet{Bertelli_2006}.

Moreover, we have shown that the method based on the filling rate of
$D$ reduces the number of operations between 19 \% and 32 \%, as
compared to the spiral approach and between 30 \% and 50 \%, as
compared to the naive method, on standard images. Even on ``random''
image the improvements is of 3 \% (resp 18 \%) compared to the
spiral (resp. naive) method.

Consequently, the filling rate of the distances array, combined with
the geodesic extrema when several points have the minimum filling
rate, seems to be the best criterion to fill efficiently the
distances array.

\begin{figure}[!htb]
    \centering
  \begin{tabular}{cc}
    \includegraphics[width=0.4\columnwidth]{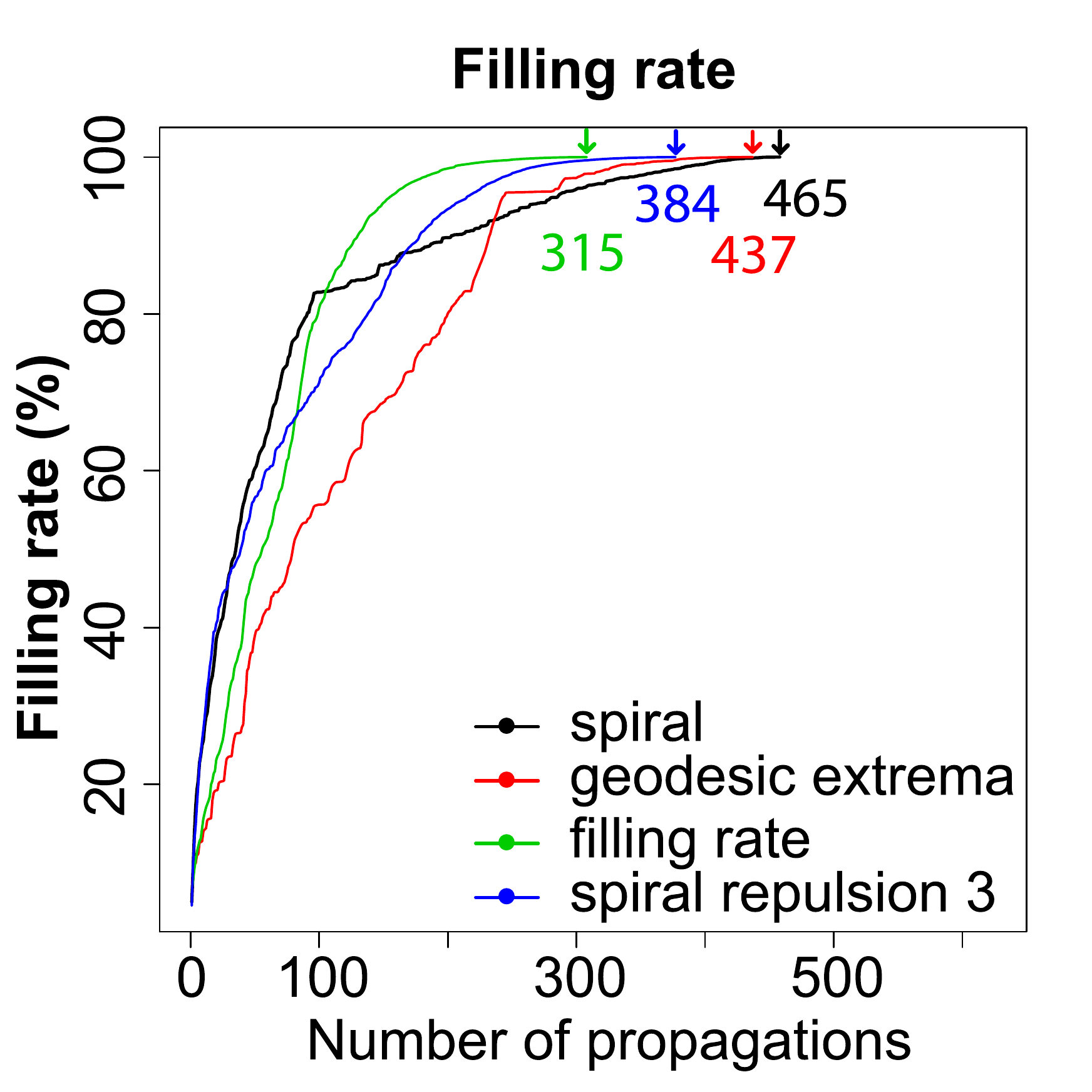}&
    \includegraphics[width=0.4\columnwidth]{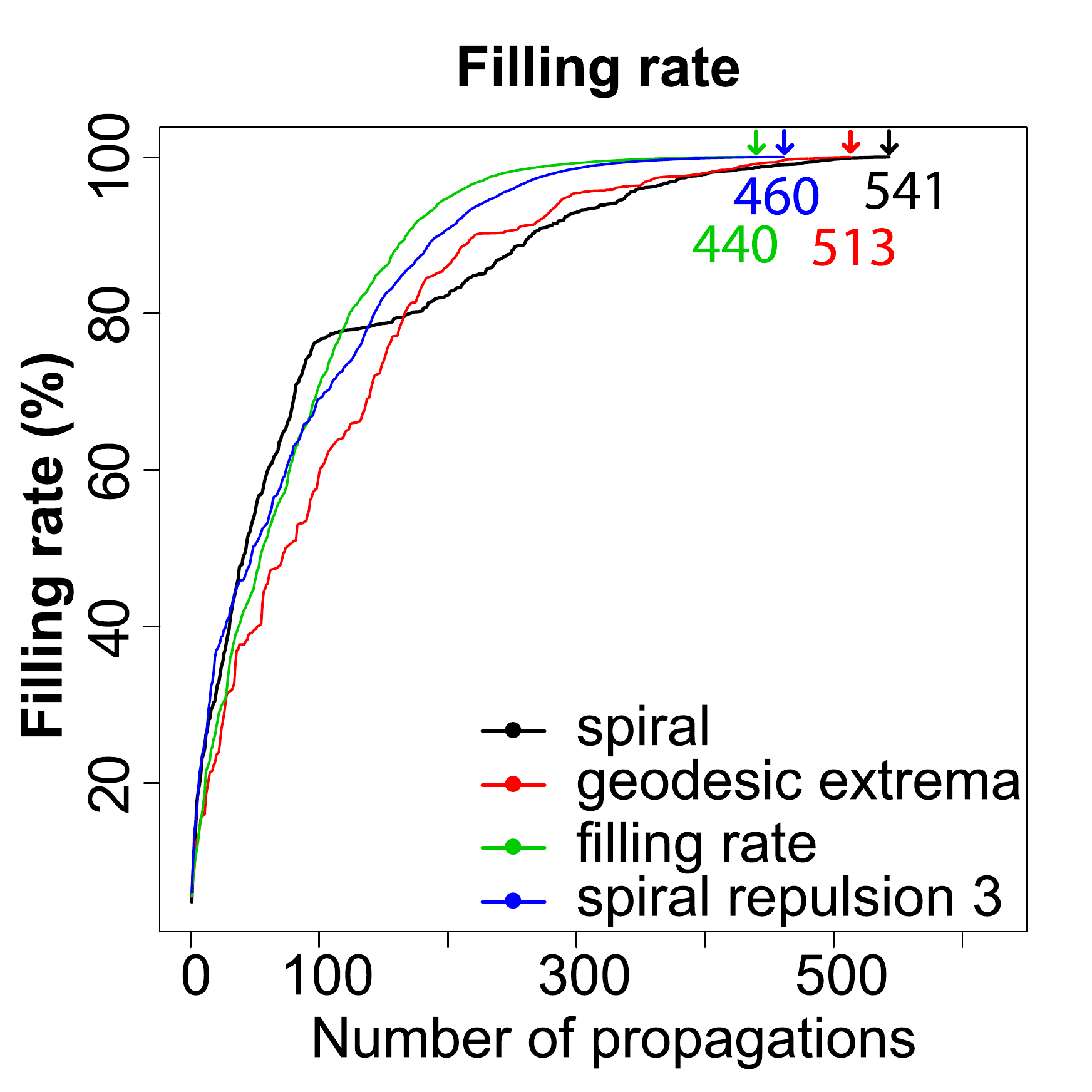}\\
    \footnotesize (a) ``bumps'' & \footnotesize (b) ``hairpin bend''\\
    \includegraphics[width=0.4\columnwidth]{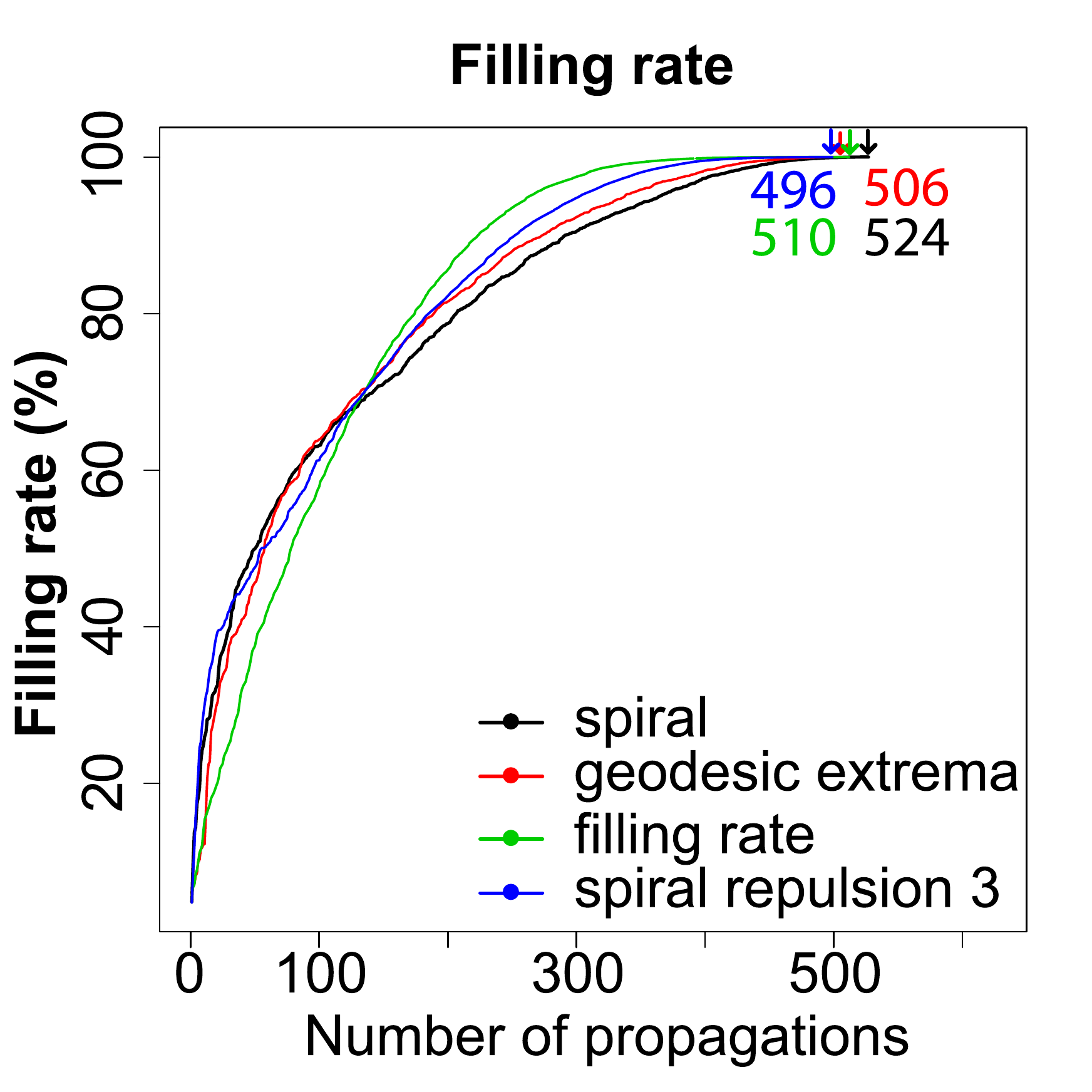}\\
    \footnotesize (c) ``random''\\
  \end{tabular}
  \caption{Comparison of the filling rates, of the array of distances,
  for the methods: spiral, geodesic extrema, the method based on the filling
  and the spiral method with a repulsion distance of 3 pixels
  for the images ``bumps'', ``hairpin bend'' and ``random''.}
  \label{fig:comparaison:comparaison}
\end{figure}

\columnbreak

\begin{table}
  \centering
  \begin{tabular}{c|c|c|c|c}

                                     & \footnotesize Spiral  & \footnotesize spiral       & \footnotesize geodesic  & \footnotesize filling \\
    \footnotesize $\Delta_r(naive)$  & \footnotesize method  & \footnotesize method with  & \footnotesize extrema   & \footnotesize rate\\
                                     &                       & \footnotesize repulsion    & \footnotesize method    & \footnotesize method  \\
    \hline
    \footnotesize ``Bumps''        & \footnotesize 25.6 \%  & \footnotesize 38.6 \%          & \footnotesize 30.1 \% & \footnotesize \textbf{49.6 \%}\\
    \footnotesize ``Hairpin bend'' & \footnotesize 13.4 \%  & \footnotesize 26.4 \%          & \footnotesize 17.9 \% & \footnotesize \textbf{29.6 \%}\\
    \footnotesize ``Random''       & \footnotesize 16.2 \%  & \footnotesize \textbf{20.6 \%} & \footnotesize 19.0 \% & \footnotesize 18.4 \%\\

  \end{tabular}\\ \vspace{0.2cm}
  (a)\\
    \vspace{0.5cm}
    \begin{tabular}{c|c|c|c}

                                     & \footnotesize spiral       & \footnotesize geodesic  & \footnotesize filling \\
    \footnotesize $\Delta_r(spiral)$  & \footnotesize method with  & \footnotesize extrema   & \footnotesize rate\\
                                     & \footnotesize repulsion    & \footnotesize method    & \footnotesize method  \\
    \hline
    \footnotesize ``Bumps''        & \footnotesize 17.4 \%    & \footnotesize 6.0 \%       & \footnotesize \textbf{32.3 \%} \\
    \footnotesize ``Hairpin bend'' & \footnotesize 15.0 \%    & \footnotesize 5.2 \%       & \footnotesize \textbf{18.7 \%} \\
    \footnotesize ``Random''       & \footnotesize  \textbf{5.3 \%} & \footnotesize 3.4 \% & \footnotesize  2.7 \%\\

  \end{tabular}\\ \vspace{0.2cm}
  (b)
  \caption{ Relative difference values of filling rates \emph{(a)} between the different methods and the naive approach,
  or \emph{(b)} between the different methods and the spiral approach.
  For each image is given in bold the best relative difference value.}
  \label{tab:comparaison:comparaison}
\end{table}

\section{Conclusion}

From a comparison between different approaches, it turns out that a
method based on the optimization of the filling rate of the
distances array is the most efficient to compute the geodesic
distances between all pairs of pixels in an image.

Besides, in the current paper, we have shown our results on grey
images. They can be directly extended to hyperspectral images using
appropriate pseudo-metrics \citep{Noyel_IAS_2007}. This can be a
useful step for a subsequent clustering by kernel methods or data
reduction approaches on multivariate images.

The main motivation for our developments is the computation of all
pairs of geodesic distances for the pixels of an image, which is
usually a graph of thousands of vertices arranged spatially.
Nevertheless, our approach is valid on more general graphs, than
those associated to bitmap images, after determining the ``boundary
vertices'' of the graph, since then, the computation of the geodesic
centre (and the geodesic extremities) of a graph can be obtained by
a first propagation from the ``boundary vertices''. The ``boundary
vertices'' can be defined, for instance, as the vertices having less
neighbouring vertices than the average number of connectivity in the
graph.

\columnbreak
\bibliography{refs}
\end{paper}
\end{document}